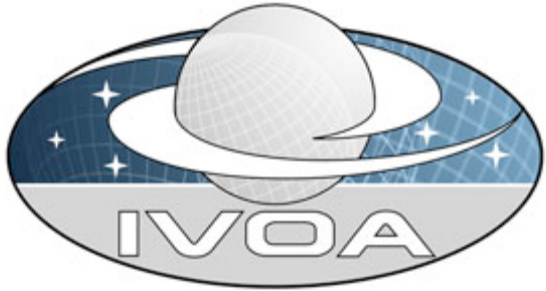

*I*nternational

*V*irtual

*O*bservatory

*A*lliance

# Simple Spectral Access Protocol

# Version 1.1

## IVOA Recommendation, February 10 2012

**This version:**
 http://www.ivoa.net/Documents/SSA

**Latest version:**
 http://www.ivoa.net/Documents/latest/SSA.html

**Previous version(s):**
 Version 1.1 PR, http://www.ivoa.net/Documents/SSA/20111018
 Version 1.1 PR, http://www.ivoa.net/Documents/SSA/20110706
 Version 1.1 PR, http://www.ivoa.net/Documents/SSA/20110417
 Version 1.04, February 2008
 Version 1.03, December 2007
 Version 1.02, September 2007
 Version 1.01, June 2007
 Version 1.00, May 2007
 Version 0.97, November 2006
 Version 0.96, September 2006
 Version 0.95 May 2006
 Version 0.91 October 2005
 Version 0.90 May 2005

**Editor:**
 D.Tody

**Authors:**
 D.Tody, M. Dolensky, J. McDowell, F. Bonnarel, T.Budavari, I.Busko, A. Micol, P.Osuna, J.Salgado, P.Skoda, R.Thompson, F.Valdes, and the data access layer working group.

## Abstract

The Simple Spectral Access (SSA) Protocol (SSAP) defines a uniform interface to remotely discover and access one-dimensional spectra.  SSA is a member of an integrated family of data access interfaces altogether comprising the Data Access Layer (DAL) of the IVOA. SSA is based on a more general data model capable of describing most tabular spectrophotometric data, including time series and spectral energy distributions (SEDs) as well as 1-D spectra; however the scope of the SSA interface as specified in this document is limited to simple 1-D spectra, including simple aggregations of 1-D spectra.

The form of the SSA interface is similar to that of the older Simple Image Access (SIA) interface for accessing 2-D image data, and the cone search interface for accessing astronomical catalogs. Clients first query the global resource *registry* to find services of interest. Clients then issue a *data discovery query* to selected services to determine what relevant data is available from each service; the candidate datasets available are described uniformly in a *VOTable* format document which is returned in response to the query. Finally, the client may *retrieve selected datasets* for analysis.

Spectrum datasets returned by an SSA spectrum service may be either precomputed, archival datasets, or they may be *virtual data* which is computed on the fly to respond to a client request. Spectrum datasets may conform to a standard *data model* defined by SSA, or may be *native* spectra with custom project-defined content. Spectra may be returned in any of a number of standard *data formats*. Spectral data is generally stored externally to the VO in a format specific to each spectral data collection; currently there is no standard way to represent astronomical spectra, and virtually every project does it differently. Hence spectra may be actively *mediated* to the standard SSA-defined data model at access time by the service, so that client analysis programs do not have to be familiar with the idiosyncratic details of each data collection to be accessed. Services are self describing, and provide a *service metadata query* operation which may be called to determine the capabilities of a specific service instance. Metadata returned by a service metadata query may be cached in the registry to facilitate registry-based service discovery.

Since SSA is part of a family of interfaces, much of the SSA interface described herein is common with the other DAL interfaces and not specific to SSA. In particular, the HTTP-based basic service profile, the main query parameters, and most of the dataset metadata returned in the query response, are generic and apply equally well to any type of data, and are (or will be, as interfaces are updated) shared by all the DAL interfaces.

## Status of This Document

This document has been produced by the IVOA **Data Access Layer** Working Group. It has been reviewed by IVOA Members and other interested parties, and has been endorsed by the IVOA Executive Committee as an IVOA Recommendation. It is a stable document and may be used as reference material or cited as a normative reference from another document. IVOA's role in making the Recommendation is to draw attention to the specification and to promote its widespread deployment. This enhances the functionality and interoperability inside the Astronomical Community.

This document is an update to the earlier SSA V1.04 IVOA Recommendation (February 2008). Changes from the earlier version are summarized in section 9.1.

Note from the V1.0 document (this is still planned for a future version): a *getCapabilities* operation returning service metadata will be added which will eventually obsolete the current `FORMAT=METADATA` mechanism. As an addition to the interface, this change is expected to be backwards compatible with existing services. The *getCapabilities* operation will be compatible with the VO Support Interfaces (VOSI) specification of the IVOA Grid & Web Services working group (GWS 2011). Additional changes are expected when other Grid and query language technology is integrated into the DAL interfaces including SSA.

A list of current IVOA Recommendations and other technical documents can be found at http://www.ivoa.net/Documents/.

# Acknowledgements

This document has been developed with support from the 5[th] and 6th Framework Programmes of the European Community for research, technological development and demonstration activities, contracts HPRI-CT-2001-50030, VOTech-011892, and via a grant from the National Science Foundation's Information Technology Research program to develop the U.S. National Virtual Observatory.

Many of the ideas in this document originated from others involved in developing Virtual Observatory concepts and standards. In particular, the idea of using association in the query response to group similar datasets grew out of an idea originally proposed by Roy Williams. Arnold Rots originated the idea of ranking query results via a *score* heuristic, and helped put the coordinate systems used in SSA on firm theoretical foundation via the development of STC. Francois Bonnarel, Mireille Louys, Alberto Micol, and others contributed to the representation of astronomical metadata and in particular the Characterization data model. Laszlo Dobos contributed early implementations of the access protocol using the spectral archive at JHU.

Many thanks to all who contributed to the DAL survey among spectral data providers and consumers (Dolensky/Tody 2004): Ivo Busko, Mike Fitzpatrick, Satoshi Honda, Stephen Kent, Tom McGlynn, Pedro Osuna Alcalaya, Benoît Pirenne, Raymond Plante, Phillipe Prugniel, Enrique Solano, Alex Szalay, Francisco Valdes and Andreas Wicenec.

Parts of this protocol were adapted from the OpenGIS (Open Geospatial Consortium, Inc.) Web-Mapping Service (WMS) specification. In particular, the basic service elements and certain details of the use of the HTTP protocol to formulate requests and responses is patterned after the OpenGIS WMS service. Parts of the text of this specification were adapted directly from the WMS service specification.# Contents

# 1  Introduction

The Simple Spectral Access protocol (SSAP, SSA) defines a uniform interface to remotely discover and access simple 1-D spectra.  Basic usage is similar to the Simple Image Access (SIA) protocol (Tody/Plante 2004) and the simple cone search (SCS) protocol for simple access to astronomical catalogs.  Unlike these earlier interfaces, spectral data access via SSA may involve active transformation of data as stored externally into a standard data format and data model defined by SSA, in order to deal with the problem of heterogeneous spectral data formats as stored externally.  SSA also defines much more complete metadata to describe the available datasets.

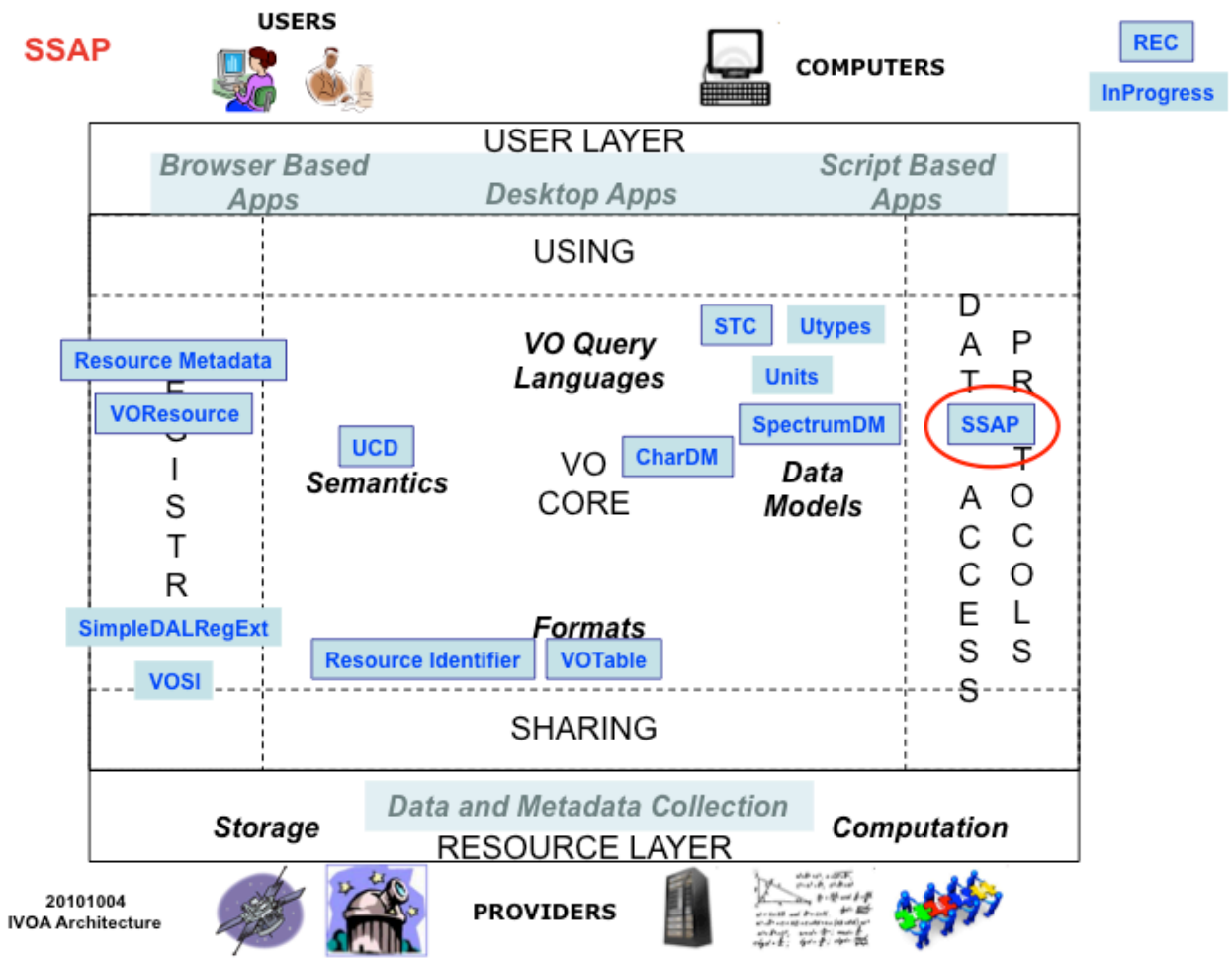

**Figure 1:**  Role of SSA within the overall IVOA architecture.

SSA is based on the Spectrum Data Model (itself making reference to the Characterisation data model) that is capable of describing most tabular spectrophotometric data, including time series and spectral energy distributions (SEDs) as well as 1-D spectra. SSAP can be used from VO applications to access the associated spectrum resources. As with most of the IVOA Data Access Protocols, SSAP makes use of VOTable for metadata exchange, STC, UCD, Utypes and Units for metadata description. An SSAP service is to be registered into the VO Registry, using the associated registry standards (in particular Resource Metadata, VOResource and SimpleDALRegExt specifically for Data Access Protocols such as SSAP). Once registered, an SSAP service will get a unique IVOA Resource Identifier. Furthermore, an SSAP service should be registered with its supported interfaces through the VOSI standard.

## 1.1 Architecture

All the IVOA data access interfaces share the same basic interface, differing mainly in the type of data being accessed. A query is used for data discovery, and to negotiate with the service the details of the *static* or *virtual* (dynamically created) datasets to be retrieved. Subsequent data access requests can then be made to retrieve individual datasets of interest. SSA differs from some other data access interfaces in that a service may **mediate** not only dataset metadata, but the actual dataset itself, to allow a client to do detailed analysis on a spectrum without having to understand how it is represented externally. Direct access to native project data is also provided.

All of the second generation DAL interfaces share the same basic service profile, although services may define additional operations specific to the service. A single service may support multiple **operations** or *methods* to perform various functions. The current DAL interfaces use an HTTP GET-based interface to submit parameterized requests, with responses being returned as structured documents, e.g., FITS or VOTable. The service operations currently defined for SSA are the following:

1. A **queryData** operation returns a table (VOTable format) describing candidate datasets which can be retrieved, including standard metadata describing each dataset, and an access reference which can be used to retrieve the data. The queryData operation is used both for data discovery and for negotiation with the service of the details of any virtual datasets to be generated.

2. A **getData** operation (currently specified in SSA merely as a data access URL) is used to access an individual dataset. The accessed data may be generated on-the-fly by the service at access time, e.g., to reformat or subset the data.

3. A metadata service call returns a table in VOTable format describing supported input and output parameters as defined in section 6 (for compatibility with earlier DAL interfaces this is not defined as a formal service operation, but as a variant of the queryData operation).

Further operations are planned but not currently defined (3.3).

A spectrum conforming to the SSA (Spectrum) data model may be returned serialized in any of a number of different data formats, including VOTable, FITS binary table, and native

XML. Comma or tab separated value (CSV) format may also be provided by implementations, but is not currently specified.

## 1.2 Basic Usage

Although SSA is a complex interface, the most common usage can be quite simple. A query can be entered in a Web browser, viewing the results as XML in the browser and downloading selected spectra by a copy-paste operation on the given access reference URL. A simple query might search for 1-D spectra by position on the sky – the classic "cone search" type of query. More complex queries are little more complicated, merely adding additional query constraint parameters, e.g., to constrain the waveband or spectral resolution, or to find spectra by redshift.

In a simple case of a positional query the SSA query URL is very similar to that for SIA or SCS. For example,

> **Example:**
> ```
> http://www.myvo.org/ssa?REQUEST=queryData&POS=22.438,-17.2&SIZE=0.02
> ```

The query response is a VOTable describing each candidate dataset as defined later in this document.

Dataset retrieval is then a simple matter of examining the query response, selecting the dataset or datasets to be retrieved, if any, and retrieving them by reading the document pointed to by the **access reference** (a URL) for the dataset. Interpretation of the returned spectrum dataset is the responsibility of the client application.

For a fully compliant SSA service, the data returned by the service will be in one of the SSA defined standard data formats, conformant to the SSA-defined Spectrum data model. Due to the need to mediate external data or support features such as format conversion or data subsetting, the service may compute the output dataset on demand, however this is transparent to the client.

## 1.3 Basic Service Elements

The basic form of a SSA service (or any other second generation DAL service) is specified in detail in section 8. In the current section we merely summarize the basic elements of a standard data service.

### 1.3.1 Request Format

In general a service may implement multiple **operations**, such as queryData; altogether these define the **interface** to the service. Interfaces may change with time hence are versioned. It is possible for a given service instance to simultaneously expose multiple interfaces or versions of interfaces.

The SSA interface described in this document is based on a distributed computing platform (DCP) comprising Internet hosts that support the Hypertext Transfer Protocol (HTTP). Thus, the online representation of each operation supported by a service is composed as a HTTP Uniform Resource Locator (URL).

A request URL is formed by concatenating a **baseURL** with zero or more operation-defined **request parameters**. The baseURL defines the network address to which request messages are to be sent for a particular operation of a particular service instance on a particular server. Service operations generally share the same baseURL but this is not required.

### 1.3.2 Parameters

Parameters may appear in any order. If the same parameter appears multiple times in a request the operation is *undefined* (if alternate values for a parameter are desired the *range-list* syntax may be used instead). Parameter *names* are case-insensitive. Parameter *values* are case-sensitive unless defined otherwise in the description of an individual parameter.

All operations define the following standard parameters:

REQUEST     The request or operation name, e.g., "queryData" (mandatory).

VERSION      The version number of the interface (optional).

The values of both the REQUEST and VERSION parameters are case-insensitive. Although the SSA V1.0 only defines a single queryData operation, use of REQUEST is mandatory to provide upwards compatibility with future versions.

A given service instance may support multiple versions of the SSA interface, which includes both the input parameters and the query response with all of its complex metadata, and by default the service assumes the highest *standard* version which is implemented (access to any experimental versions supported by a service requires explicit specification of the version by the client). Explicit specification of the interface version assumed by the client is necessary to ensure against a runtime version mismatch, e.g., if the client caches the service endpoint but a newer version of the service is subsequently deployed. If desired the client can omit the VERSION parameter to disable runtime version checking, and default to the highest version standard interface implemented by the service.

All other request parameters are defined separately for each operation.

### 1.3.3 Parameter Values

Integer numbers are represented as defined in the specification of integers in XML Schema Datatypes. Real numbers are represented as specified for double precision numbers in XML Schema Datatypes. Sexagesimal formatting is not permitted, either for parameter input or in output metadata, other than in ISO 8601 formatted time strings (sexagesimal format is fine for a user interface but inappropriate for a lower level machine interface, where it only complicates things).

SSA defines a special **range-list** format for specifying numerical ranges or lists of ranges as parameter values. For example, "`1E-7/3E-6;source`" could specify a spectral bandpass defined in the rest frame of the source. The syntax supports both open and closed ranges. Ranges or range lists are permitted only when explicitly indicated in the definition of an individual parameter. For a full description of range list syntax refer to section 8.7.2.

### 1.3.4 Error Response

In the case of an error, service operations should return a VOTable containing an INFO element with name `QUERY_STATUS` and the value set to "`ERROR`". More fundamental service or protocol errors may however result in an HTTP level error, hence a client program should be prepared to handle either response. A null query, that is a queryData which does not find any data, is not considered an error. More information on error responses is given in section 8.10.

## *1.4 Requirements for Compliance*

The keywords "**must**", "**required**", "**should**", and "**may**" as used in this document are to be interpreted as described in the W3C specifications (IETF RFC 2119). Mandatory interface elements are indicated as **must**, recommended interface elements as **should**, and optional interface elements as **may** or simply "may" without the bold face font.

Sometimes the extent to which a given interface element is required depends upon the mode of operation of the service. For example, a service which performs spectral extraction **must** implement the APERTURE query parameter, but it is not used for other types of SSA services, and for these need not be implemented.

### 1.4.1 Levels of Compliance

In order to be *minimally compliant* a service must implement all elements of the SSA protocol identified as "must" in this document. In brief, the minimal service implementation includes the following:

1. The SSA query method **must** implement the HTTP GET interface, returning the query response encoded as a VOTable document. At least the POS, SIZE, TIME, BAND, and FORMAT query parameters **must** be supported by the service (regardless of whether these are defined for the data being accessed). The query response **must** include all metadata fields identified as mandatory in the protocol.

2. The direct URL-based *getData* method **must** be provided capable of returning data in at least one of the SSA-compliant data formats (VOTable is suggested if only one format is supported).

3. The "FORMAT=METADATA" metadata query feature **must** be provided to return service metadata encoded as defined herein.

If a service cannot return data which is SSA (i.e., Spectrum DM) compliant, it is still useful to implement a service which provides a SSA-compliant query method but which returns native or external data. Such a service is said to be *query compliant* if the query operation is at least minimally compliant. The ability to return native project data is always desirable, as this provides the maximum transfer of information from the project, however the ability to return SSA (Spectrum DM) compliant data is essential for transparent multiwavelength data analysis, hence is the primary requirement. Legacy data providers are encouraged to both provide data in both their proprietary legacy data format as well as in the Spectrum DM format, leaving the choice of which is more useful for analysis up to the client application and the user.

A service is said to be *fully compliant* if, in addition to the functionality required to be minimally compliant, the service implements all the "should" elements of the interface defined herein.

A top of the line service will be fully compliant plus will implement some of the optional ("may" provide) elements of the interface. For example the service may support additional query parameters or may return additional metadata; the service may provide access to native data as well as SSA-compliant data, or may be capable of returning data in any supported standard data format requested by the client.

# 2 Concepts and Terminology

## 2.1 Dataset and Data Collection

The term **dataset** as used in this specification normally refers to a **primary dataset** such as an individual spectrum, image, table, and so forth, i.e., an individual data object usually including associated metadata. A **complex dataset** is some logical association or aggregation of primary datasets, often of different types, possibly with additional high level metadata describing the association. In common usage, *dataset* can refer to either of these. A **data collection** is a collection of primary or complex datasets, such as a survey data release (e.g., "SDSS DR6") or an instrumental data collection from an individual observatory instrument.

## 2.2 Data Model

SSA consists of both an access protocol and interface, and an underlying data model describing the data to be accessed. The term **data model** as used here refers to a logical model for the data detailing the decomposition of a complex dataset into simpler elements, including specifying the meaning of each element, the relationships between elements, the metadata used to describe the data elements and the overall dataset, and the concepts upon which the data model is based. In this document we refer to the underlying data model interchangeably as the SSA data model or the **Spectrum or spectral data model (SDM).** The data model used in SSA is described in (McDowell, Tody, et.al, 2007).

The spectral data model is an abstract model describing spectrophotometric data. The SSA protocol is an access interface which uses the SDM in a particular context, i.e., for discovery of and access to spectral data. In order to simplify usage, e.g., when the same query is issued to many different SSA services, SSA may constrain the underlying model, for example by fixing the allowable units for a time interval or spectral bandpass. SSA may also extend the underlying model by defining additional metadata specific to the access protocol itself. Hence in cases where details of the data model differ between SSA and the SDM, the SSA specification takes precedence. For any details of the data model not defined by SSA one should refer to the SDM specification, which is the primary reference for the underlying spectral data model.

Explicitly defining the data model assumed by a data object is important for a variety of reasons. Doing so helps greatly to document the structure and meaning of the data. Data analysis software has to understand data at a fundamental level in order to function correctly.

Data model **mediation** - the process of transforming data from some externally-defined data model to a prescribed data model (the SSA data model in our case) - makes it possible for a client application to deal uniformly with external data without having to understand the idiosyncratic details of each external data collection. SSA does data model mediation on the fly, at data access time, in the service used to publish a data collection to the VO. A data publishing service is written for a specific data collection by the creators or curators of the data who understand the data well, and may thereafter be accessed by any number of independently written client applications; hence mediation to a standard model is best performed by the service.

If more detailed knowledge of a specific data collection is required than is possible using a standard model, direct pass-through of the *native project data* is also possible. This is an important capability as it ensures that nothing has been lost in the translation, and it provides for direct communication between the client application (or user) and the data provider. Nonetheless, for general automated multiwavelength data analysis, if we provide only access to native project data, this puts the burden of interpretation of individual project datasets completely on the data consumer (e.g., the client application), and we feel that the *data provider* has a better understanding of their data, and is generally better equipped to make this translation. Hence data should always be provided in a form compliant to the SSA/Spectrum data model if possible, with pass-through of native project data provided as well where possible.

## 2.3 Data Representation

A data model defines the logical content of data, but says nothing about how the data is represented externally. The same data object may be represented externally in many different ways, e.g., as a FITS file or VOTable, as a direct XML serialization, in a RDBMS, and so forth. So long as the data model does not change, and the data representation is expressive enough, data may be transformed from one representation to another without loss of information. If transformation between different data models is required, some loss of information may occur. This can happen, for example, during mediation of external data to a known data model by a SSA service.

In the most general case SSA uses a container-component approach to represent datasets. In this case a general container such as VOTable or FITS is used to represent a Spectrum object. A similar approach is used for the SSA query response, which is returned as a VOTable. The container is used to aggregate **component data models** which are associated in some fashion to model more complex objects such as a spectrum. The advantage of this approach is flexibility, in that there is no fixed structure for the overall dataset, and extensibility, as it is easy to add custom components to describe the details of a specific data collection while conforming to the standard core model.

Application programs typically manipulate a data object by directly accessing the elements of the data model via some language-specific API. **UTYPE** tags are used to provide a uniform means to identify the elements of a data model in any language or environment. For example, given the component data model "DataID", the UTYPE "DataID.Title" identifies the data model field containing the title string for the dataset; "DataID.Collection" identifies the parent data collection, and so forth.

## 2.4 Virtual Data

A **virtual dataset** is one which can be described, but which may not physically exist until it is accessed, at which time it is created on the fly by the service. A typical example is a cutout (subset) of an image or spectrum. Where general distributed multiwavelength data analysis is concerned, most data access in the VO is necessarily to virtual data. Physical datasets can also be accessed, but this is a far less powerful technique as physical datasets are often too large to transmit efficiently over the network, particularly when only a small portion of the data is needed, and capabilities such as mediation to a standard model or transformations of various kinds are not possible.

When a query is made to a SSA service which can return virtual data, the service computes the parameters of any virtual datasets it can generate to satisfy the query. What can be generated depends upon what the client has requested, the input data available to the service, and the capabilities of the service. The metadata returned in the query response will describe the virtual dataset and its relationship to any parent dataset or datasets. The access reference is in effect a token to be passed back to the service to generate the virtual dataset. The client can either access the virtual data (in which case it is realized by the service, and returned), or further refine the query to more finely specify the data to be returned by the service.

## 2.5 Data Derivation

Data can come from a variety of sources, and may go through various types of processing, including by the data access service itself, before being delivered to a client analysis application. It is important for analysis to understand the origin of the data and what processing it has undergone. To address this issue we introduce two new concepts, *data source* and *creation type*.

### 2.5.1 Data Source

The data source specifies where the data originally came from, that is, the data collection to which the service provides access. The following values are currently defined:

| survey | A survey dataset, which typically covers some region of observational parameter space in a uniform fashion, with as complete as possible coverage in the region of parameter space observed. |
|--------|---|
| pointed | A pointed observation of a particular astronomical object or field. Typically these are instrumental observations taken as part of some PI observing program. The data quality and characteristics may be variable, but the observations of a particular object or field may be more extensive than for a survey. |
| custom | Data which has been custom processed, e.g., as part of a specific research project. |
| theory | Theory data, or any data generated from a theoretical model, for example a synthetic spectrum. |
| artificial | Artificial or simulated data. This is similar to theory data but need not be based on a physical model, and is often used for testing purposes. |

### 2.5.2 Creation Type

The creation type describes the process used to produce the dataset as returned by the service, from the data source. Typically this describes only the processing performed by the data service, but it could describe some additional earlier processing as well, e.g., if data is partially precomputed. The creation type is especially important for virtual data and for data which is derived from the parent data set by some complex form of processing. The following values are currently defined:

| archival | The entire archival or project dataset is returned. Transformations such as metadata or data model mediation or format conversions may take place, but the content of the dataset is not substantially modified (e.g., all the data is |
|---|---|

| | returned and the sample values are not modified). |
|---|---|
| cutout | The dataset is subsetted in some region of parameter space to produce a subset dataset. Sample values are not modified, e.g., cutouts could be recombined to reconstitute the original dataset. |
| filtered | The data is filtered in some fashion to exclude or alter portions of the dataset, e.g., passing only data in selected regions along a measurement axis, or processing the data in a way which recomputes the sample values, e.g., due to interpolation or flux transformation. Filtering is often combined with other forms of processing, e.g., projection. |
| mosaic | Data from multiple non- or partially-overlapping datasets are combined to produce a new dataset. |
| projection | Data is geometrically warped or dimensionally reduced by projecting through a multidimensional dataset. |
| spectralExtraction | Extraction of a spectrum from another dataset, e.g., extraction of a spectrum from a spectral data cube through a simulated aperture. |
| catalogExtraction | Extraction of a catalog of some form from another dataset, e.g., extraction of a source catalog from an image, or extraction of a line list catalog from a spectrum (not valid for a SSA service). |

The full creation type may involve more than one of these operations, for example, both projection and filtered, or both spectral extraction and filtered.

This list is by no means complete in general astronomical data processing terms, but is intended to express only the types of operations which might take place during VO data access, where subsetting, filtering, projection, spectral extraction, etc., are all defined operations. Other values may be added in the future. The creation type is not intended to describe the processing done to produce the data collection itself, which the service is used to access.

## 2.6 Service Type

Not all SSA services are of the same type: services are further classified by their subtype, indicating how they generate the spectra returned by the service. The subtype of a SSA service is similar to the dataset creation type as described in section 2.5.2; usually the creation type and the SSA service subtype are the same, but this is not always the case. A simple service which returns only entire archival spectra is an "*archival*" SSA service. A service which can return subregions of larger spectra is a "*cutout*" service. A SSA service which can combine multiple input spectra is a "*mosaic*" service (a mosaic service can also do cutouts if presented with a sufficiently small spectral bandpass). A SSA service which dynamically generates spectra from more fundamental data, e.g., a spectral data cube or event list, is a "*spectralExtraction*" service.

## 2.7 Services, Interfaces, and Protocols

A **service** operates at a defined **service endpoint** (e.g., an Internet URL, often called a baseURL), and implements one or more predefined client-server **interfaces**. The service interface consists of one or more **service operations**, also known as requests, or methods. Each operation accepts as input zero or more **request parameters**. The details of how a client talks to a service interface over a given transport protocol (e.g., HTTP) defines the **protocol** used to interact with the service.

## 2.8 Dataset Identifiers

A **dataset identifier** is an identifying name for a dataset that is globally unique within the VO and is compliant with the URI syntax rules (IETF RFC 2396). It consists of an **IVOA Identifer** (Plante et.al. 2005), followed by a pound sign ("#"), and a local identifier. The IVOA Identifier defines a name space (for example a data collection) which may contain any number of individual datasets, each with its own unique local identifier. The local identifier consists of one or more legal URI characters, and is a name given by the creator or publisher of the dataset which identifies an individual dataset within the namespace defined by the IVOA Identifier..

In ABNF (IETF RFC 2234) format, the dataset identifier is defined as:

> *dataset-id = ivoa-id "#" uric*

where *ivoa-id* is a legal IVOA identifier in URI format (uri-form in [Identifiers]) and *uric* is the set of legal URI characters (*uric* in (IETF RFC 2396)).

To provide consistency with the IVOA Identifier standard, the rules for comparing dataset identifiers are the same as for IVOA identifiers: two dataset identifiers shall be considered as refering to the same dataset "if a case-insensitive, character-by-character comparison indicates that they are identical." That is, "apart from a transformation to handle case-insensitive comparisons, no other normalizing transformations shall be necessary" to test for equivalence [Identifiers].

As we shall see in section 4.2.5.5, we define several types of dataset identifiers, including **CreatorDID**, **PublisherDID**, and **DatasetID**. The CreatorDID is the dataset identifier (if any) assigned by the creator of the dataset, for example a survey project or observatory. This does not change, even if the dataset is published in multiple locations. CreatorDIDs can be assigned at dataset creation time, before the data has been published to the VO, but will be globally unique so long as the creating entity uses a registered IVOA Identifier for the namespace. The PublisherDID is the dataset identifier assigned by a publisher; this DID is unique within the publisher's name space, but has no meaning otherwise. A special case of a PublisherDID is a DatasetID, which is a globally unique dataset identifier assigned by a publisher to attempt to index data from many sources, for example an ADS dataset identifier.

When data is published to the VO it should always be possible for the publisher to assign a unique PublisherDID. A CreatorDID may or may not be assigned by the dataset creator (legacy data at least is unlikely to have one). We recommend the practice as it can easily be done in an automated fashion at dataset creation time, as one might assign a serial number, and provides a globally unique way to identify any dataset. In general a global data indexing service will only index selected datasets, e.g., those referenced in journal articles, so while a DatasetID can be useful for things such as linking datasets to journal articles, many datasets may not have registered DatasetIDs, and in principle there can be multiple publishing authorities registering DatasetIDs.

## 2.9 Provenance

The combination of a data source with a creation type provides us with a primitive capability for describing the ***provenance*** of a dataset, i.e., where it came from, and how it was produced. This is important because SSA and other DAL services can generate virtual data products where complex processing may be performed at access time.

To be able to describe the provenance of a virtual data product we need one additional concept, the dataset identifier of the parent dataset, as assigned by the entity which created the dataset (typically a survey project, observatory, modeling program, etc.). Dataset identifiers are discussed in more detail in section 2.8.

Given a virtual data product we can then say how the data product was derived from the parent dataset or datasets (the creation type), identify the parent dataset (the creator-assigned dataset ID), and the origin and type of data from which the virtual data product was derived (data source, collection, and so forth). In the more complex cases such as a mosaic a virtual data product may have multiple parent datasets.

If a process which produces data products is complex enough, with many inputs, ultimately the result is a new data collection, but in most runtime data access scenarios the simple provenance model presented here should be enough to identify a virtual data product or other dataset and how it was produced.

## 2.10 Data Association

There are many cases where it is desirable to be able to associate multiple datasets, for example to model a multi-spectral observation such as an Echelle, or to group datasets that represent the same data made available in several different data formats. Spectra of the member galaxies in a cluster might be a completely different type of association. In the case of images, a multi-band observation could be viewed as an association of several independent images, each in a single spectral band and with some shared observational metadata.

The approach taken in SSA to address this problem of ***complex data*** is to keep the basic data objects as simple as possible but use ***association*** to describe more complex entities. Hence, an Echelle observation could be viewed as a collection of independently accessible 1-D spectra which are logically associated. The spectra would include the individual Echelle orders and possibly an overall combined high resolution spectrum. Some extension metadata might also be provided to provide additional information describing the overall association. The individual spectra would be usefully accessible without requiring that a client application understand the complex instrument (an Echelle spectrograph) which produced the data, however the more complex view would optionally be accessible as well.

Associations are described in the SSA query response since this has the ability to relate multiple datasets. How this is done will be described further in the specification of the SSA query response, but the main technique is to define a new query response field Association.ID for which all members of an association share the same value. An association key may also be provided for each member of the association to uniquely identify their role within the association (e.g., the Echelle order in our example above). Finally, an association Type field or param tells what type of association this is. The ID may be used to link to extension metadata providing further information describing the specific extension.

## *2.11 UTYPEs and UCDs*

A **UTYPE** is a fixed string which uniquely identifies a field of a data model irregardless of representation. UTYPEs are strings such as "Target.Name", using embedded period characters to delimit the fields of the UTYPE. A simple way to think of a UTYPE is as a reference to a field of a data structure in a language such as C. The effect is to flatten a hierarchical data model so that all fields of the data model are represented by fixed strings in a flat name space, allowing a wide variety of software to be used to manipulate or use the model. Of course if a data model becomes complex enough this will no longer be possible, but the approach has significant benefits for a wide variety of data. UTYPEs are defined within a single name space identifying the data model, and are unique only within the context of the specified data model.

Note that while a UTYPE is always a fixed string which uniquely identifies a data model element, if there are *multiple instances* of the data model in a container (name space), then multiple data elements may have the same UTYPE. For example, in a VOTable representation, multiple table FIELDs may have the same UTYPE if there are multiple instances of a component data model (e.g., Association) in the table. In this case the GROUP construct is used to separately identify the data model instances. Within each GROUP, the UTYPE values still uniquely identify the field of the data model. Multiple instances of individual table FIELDs (e.g., Curation.Reference) are also possible.

A **UCD** identifies the *semantic type* of a data value or data model element, saying what type of quantity, in a physical sense, is stored in the value. UCDs are defined globally, independent of how they are used. UCDs may be used indendently of any data model. Multiple data models may define fields which share the same UCD, or multiple fields of a single data model may share the same UCD. Since multiple fields even within a single data model may share the same UCD value, UCDs cannot be used to uniqely identify data model fields. UCDs however provide a unique capability to identify or associate similar types of fields in independent data models or data instances.

Both UTYPEs and UCDs are case-insensitive, and case should be ignored when comparing string values for equality.

# 3   SSA Operations

## *3.1 Introduction*

The operations currently defined by the SSA protocol are **queryData** (mandatory), **getData** (reserved), **stageData** (reserved), **getCapabilities** (reserved) and **getAvailability** (reserved). Of these, currently only queryData is defined as an explicit parameterized operation. GetData is currently not implemented as a service operation, rather an explicit access reference URL is used to retrieve datasets. In addition a metadata query feature (`FORMAT=METADATA`) is defined which is expected to be obsoleted by the getCapabilities operation in future versions.

The specification herein of whether support for a parameter is required, recommended, or optional refers to the service, not to the query submitted by the client. Unless otherwise specified by the operation, all parameters except `REQUEST` are optional for the client (depending upon the operation, invoking an operation with *no* parameters may however result in an invalid operation).

## *3.2 Methods & Protocols*

As with all the DAL interfaces, a SSA service may eventually define interfaces for multiple "distributed computing platforms" or transports, e.g., HTTP GET/POST and SOAP. At this time only a HTTP GET interface is defined.

If the SSA query is transmitted as an HTTP GET request then the URL to express a data query is formed like this:

> `<Service.BaseURL>?VERSION=1.0&REQUEST=queryData<¶m=value…>`

**Example:**
```
http://www.myvo.org/ssa?VERSION=1.0&REQUEST=queryData&POS=22.438,-
17.2&SIZE=0.02
```

The `Service.BaseURL` is stored in the IVOA resource registry (Hanisch et al. 2005).

## *3.3 Future Extensions*

Several operations are reserved for future revisions of this specification. These operations are not yet fully defined.

### 3.3.1 GetCapabilities

The *getCapabilities* operation is reserved for future revisions to query the service for its *service metadata* (capabilities and interface).

### 3.3.2 StageData

The *stageData* operation is used in DAL interfaces to request asynchronous generation and staging (data delivery) of one or more, possibly virtual datasets, as identified in a prior call to *queryData*. Greater flexibility in staging data, specifying where generated data is to be delivered, including support for third-party data delivery e.g. via *VOSpace*, is intended to be part of what stageData will provide. Providing this capability is not planned for SSA V1.0, but may be added in a future version.

### 3.3.3 GetAvailability

The *getAvailability* operation is used by an external client (normally the VO or grid infrastructure) to monitor the availability and health of a service.

# 4   QueryData Operation (required)

The purpose of the SSA query is to determine the availability and characterization of data satisfying certain client-specified search constraints. The result is returned encoded as a VOTable document wherein each row of the table describes one candidate dataset.

## *4.1 Input Parameters*

A simple query is defined in terms of a 4-dimensional physical parameter space:

- spatial region (for SSA an aperture on the sky defined by `POS, SIZE`)
- temporal region (`TIME`)
- spectral region (`BAND`)

A minimally-compliant SSA service **must** support at least these four parameters, plus the `FORMAT` parameter, which specifies the format in which data (spectra) are to be returned. Various other parameters specific to spectrophotometric data, as outlined later in this section, are also defined and **should** also be supported by the service if possible, to be fully compliant.

Unless otherwise specified, if the service does not support a query parameter defined by the protocol it **must** permit the parameter to be present in the query without error, even if the parameter is not actually used as a query constraint by the service. Most parameters are used as query constraints, to narrow the selection of data by the service. If a given parameter is not specified in the query or is not supported by the service, or cannot be applied to the data as the necessary dataset metadata is not available (note this is different than the case of theory data described below), a logical value of "all" is generally assumed, meaning that the parameter is not used to constrain the query. This allows a query to succeed even if it includes parameters which the service does not support, so that the same query can be submitted to multiple service instances. Since queries can be imprecise (multiple candidate datasets are described in the response) it is up to the client to analyze the returned query metadata to further refine the query.

If a service supports a parameter but the value given cannot be parsed or is otherwise illegal (as opposed to merely not matching any data) then an error response should be returned to the client. If a service does not support a parameter it is not required to parse the parameter value and report errors, i.e., it may ignore the unsupported parameter.

If a service is required to support certain input parameters, that means that the service must be prepared to use such a parameter to constrain a query. If this is not done and the service merely ignores a mandatory input parameter which the service is required to support, then it may be easy for the client to pose a query which results in an overflow of the query response or some other error condition. For example, if a client queries for data based only on the spectral bandpass and the service does not support the BAND parameter, the query may overflow or be declared invalid even though valid data is available.

Specific parameters may or may not have meaningful values for a given data collection. For example, for theory data, anything having to do with time or position on the sky may be undefined. For solar or planetary data, time is defined but the spatial position on the celestial sphere may be undefined or not meaningful. In such a case, where a specific value is specified for an attribute which is undefined or has no meaning for a given data collection, the service should respond by finding no matching data (for example a query based on POS, if cast to a broad range of services, would probably not find any matching data if posed against a service providing access to theory data). For data collections where all physical measurement parameters are meaningful, for example spectra of galactic or extragalactic astronomical targets, all parameters should be supported and used to constrain the query, even if only imprecise values of the parameters are known for a given dataset.

### 4.1.1 Mandatory Query Parameters

The following parameters **must** be implemented by a compliant service:

| Parameter | Sample value | Physical unit | Datatype |
|-----------|--------------|---------------|----------|
| POS | 52,-27.8 | degrees; defaults to ICRS | string |
| SIZE | 0.05 | degrees | double |
| BAND | 2.7E-7/0.13 | meters | string |
| TIME | 1998-05-21/1999 | ISO 8601 UTC | string |
| FORMAT | votable | - | string |

All services must support queries containing at least these five parameters, representing coverage in the fundamental physical measurement axes, and the output data format or formats desired by the client. Although services must support these parameters, this does not necessarily mean that the quantity referred to is meaningful for the class of data being queried (4.1). While a compliant service must implement these parameters and use them (if specified) to constrain queries, a valid query can be composed from any combination of parameters, and may include or omit any given parameter. If a parameter is not specified, it is not used to constrain the query. For example if POS is not specified, data from any spatial region, or data for which POS is undefined, will satisfy the query and other parameters must be used to constrain the query.

#### 4.1.1.1 POS

The center of the region of interest. The coordinate values are specified in list format (comma separated) with no embedded white space, as defined in section 8.7.2.

> Example: POS=52,-27.8

POS defaults to right-ascension and declination in decimal degrees in the ICRS coordinate system. A coordinate system reference frame **may** optionally be specified to specify a coordinate system other than ICRS. The reference frame is specified as a list format modifier, with the acceptable values as defined in the respective table of the CoordSys object in the Spectrum data model (McDowell/Tody et al. 2007), which is in turned based upon the spatial coordinate frames defined by Table 3 (standard reference frames) in STC (Rots 2007).

> Example: POS=52,-27.8;GALACTIC

Coordinates requiring more than two values are possible merely by having more than two comma-delimited values before the qualifier.

Whether or not a service supports coordinate systems other than ICRS for POS is a service-defined optional capability. It is an error if a coordinate reference frame is specified which the service does not support.

#### 4.1.1.2 SIZE

The diameter of the search region specified in decimal degrees.

> Example: SIZE=0.05

A valid query does not have to specify a SIZE parameter. If SIZE is omitted in a positional query, the service should supply a default value intended to find nearby objects which are candidates for a match to the given object position.

### 4.1.1.3 BAND

The spectral bandpass is given in *range-list* format as defined in section 8.7.2, with each list element specified either numerically as a wavelength value or range, or textually as a spectral bandpass identifier, e.g., a filter or instrumental bandpass name. A service **must** support at least one bandpass list element, which may be either a single value or (for numerical ranges) an open or closed range. Multiple element range-lists **may** be supported. If a single numerical value is specified as a range element it matches any spectrum for which the spectral coverage includes the specified value. If a two valued range is given, a dataset matches if any portion of it overlaps the given spectral region. See section 8.7.2 for a more detailed discussion of range lists.

For a numerical bandpass the units are wavelength in vacuum in units of meters (Hanisch *et.al*, 2005). The spectral rest frame **may** optionally be qualified as either "source" or "observer", specified as a range-list qualifier.

> Example: `BAND=1E-7/3E-6;source`

For most queries the precision with which the spectral bandpass is specified in the query probably does not matter very much. A rough bandpass broad enough to find all the interesting data will generally suffice; the more precise spectral bandpass specified in the query response for each spectrum can then be used to refine the query. In some cases, for example a cutout service operating upon high resolution spectra, support at the service level for specifying the spectral rest frame could be important. If the service does not support specification of the spectral frame the syntax should be permitted but may be ignored.

If a bandpass is specified as a string it is assumed to be a bandpass identifier such as a filter name or instrumental bandpass, as specified in the resource metadata [RSM, Hanisch *et.al*, 2005] for Coverage.Spectral.Bandpass. A spectral bandpass specified by name is equivalent to the corresponding numerical closed range specifying the spectral coverage of the bandpass. The service **may** support bandpass names in the BAND parameter. Since there is no standard list of filter names or instrumental bandpasses, and these can overlap, there is no apriori way to know what to call a bandpass in a query; however it is possible to learn these values in a prior query to the same service and then use them to refine the query.

If the service supports query by bandpass name but does not recognize an input bandpass name, it should return an error indicating that it does not recognize the named bandpass. If the service does not support query by bandpass name but is called in this way, it should return an error.

> Example: `BAND=J`

Bandpass names are often not useful for spectra (they are probably more useful for image or time series data) but there are cases where they are useful for spectra, for example for a velocity spectrum of a specific emission line.

### 4.1.1.4  TIME

The time coverage (epoch), specified in range-list form as defined in section 8.7.2, in a restricted subset of ISO 8601 format.  If the time system used is not specified UTC is assumed.  Allowable ISO8601 formats include the date (*e.g., yyyy-mm-dd* with the month and day fields being optional, the minimum being *yyyy*), or the date-time (*e.g., yyyy-mm-ddThh:mm:ss*).  This restriction on the allowable ISO8601 formats applies throughout this document unless otherwise specified.

The value specified may be a single value or an open or closed range.  If a single value is specified it matches any dataset for which the time coverage includes the specified value.  If a two valued range is given, a dataset matches if any portion of it overlaps the given temporal region.  An imprecise value such as *yyyy* indicates the entire period specified, e.g., 1990-2000 would match any dataset overlapping the range from the beginning of 1990 to the end of 2000.

### 4.1.1.5  FORMAT

The FORMAT parameter defines the data formats the client is interested in retrieving via a subsequent *getData* call.  The value is a comma-delimited list as defined in section 8.7.2, where each element can be any recognized MIME-type such as

```
application/x-votable+xml, application/fits, application/xml,
text/csv, text/html, image/jpeg
```

and so forth.  If finer discrimination is necessary, MIME type parameterization may be used to more finely specify any format available from a service, for example

```
FORMAT=application/fits;convention=STScI-STIS
```

might specify the native project specific format defined by the HST STIS instrument (this is merely a hypothetical example; the specification of native project MIME types is outside the scope of SSA).  Normally this should not be required as `FORMAT=native` may be used to specify the native format for a data collection, if available.

In addition to the standard MIME-type format specifications defined above, the following special shorthand values are defined:

| FORMAT | Meaning |
|---|---|
| `all` | All formats supported by the service |
| `compliant` | Any SSA data model compliant format |
| `native` | The native project specific format for a  spectrum |
| `graphic` | Any of the graphics formats: JPEG, PNG, GIF |
| `votable` | Shorthand for `application/x-votable+xml, the SSA VOTable format` |
| `fits` | Shorthand for `application/fits, the SSA-compliant FITS format` |

| `xml` | Shorthand for `application/xml, the SSA native XML serialization` |
|---|---|
| `metadata` | Reserved for returning service metadata as a VOTable |

These shorthand values all assume an SSA-compliant serialization, i.e., "`fits`" refers to the SSA (Spectrum DM) FITS serialization; "`native`" would be used to instead access a native project FITS format if available. If FORMAT is omitted, FORMAT=ALL should be assumed, and the service should describe all available formats. FORMAT values are case insensitive.

The FORMAT parameter describes the desired format of returned data. If no data is available in the specified format, a null query response should be returned indicating that no data satisfying the query is available. If data is dynamically generated the service may generate data in the format requested by the client on the fly. Note FORMAT applies only to the data; the query response itself is always returned as a VOTable.

## 4.1.2 Recommended and Optional Query Parameters

The following additional parameters **should** or **may** be implemented by a service; all the recommended parameters are required for a fully compliant service. In the table below and those following, mandatory parameters are indicated by MAN, recommended parameters by REC, and optional parameters by OPT.

| Parameter | Sample value | Unit | Req | Datatype |
|---|---|---|---|---|
| APERTURE | 0.00028 (=1") | degrees | OPT | double |
| SPECRP | 2000 | dimensionless | REC | double |
| SPATRES | 0.05 | degrees | REC | double |
| TIMERES | 31536000 (=1yr) | seconds | OPT | double |
| | | | | |
| SNR | 5.0 | dimensionless | OPT | double |
| REDSHIFT | 1.3/3.0 | dimensionless | OPT | string |
| VARAMPL | 0.77 | dimensionless | OPT | string |
| TARGETNAME | mars | | OPT | string |
| TARGETCLASS | star | | OPT | string |
| FLUXCALIB | relative | | OPT | string |
| WAVECALIB | absolute | | OPT | string |
| | | | | |
| PUBDID | ADS/col#R5983 | | REC | string |
| CREATORDID | ivo://auth/col#R1234 | | REC | string |
| COLLECTION | SDSS-DR5 | | REC | string |
| | | | | |
| TOP | 20 | dimensionless | REC | int |
| MAXREC | 5000 | | REC | string |
| MTIME | 2005-01-01/2006-01-01 | ISO 8601 | REC | string |
| COMPRESS | true | | REC | boolean |
| RUNID | | | REC | string |

The spatial, spectral, and time resolution of the data may all be used as query constraints to find data of interest. The aperture size or coverage cannot be used as a query constraint (the APERTURE parameter is used only for spectral extraction), but can be determined from the query response metadata. The spectral resolution is specified as the spectral resolving power to avoid scaling effects over a wide range of wavelength. For a spectrum the time resolution is rarely significant, but is included for completeness and so that the same query interface can eventually be used for time series data. The creator and publisher dataset identifiers and data collection name may be used to precisely specify the data to be accessed. All parameters are explained in more detail below.

### 4.1.2.1  APERTURE

The aperture parameter is used only for spectral extraction, i.e., computation of spectra derived from more fundamental data such as a spectral data cube or event list, using a synthetic aperture. The aperture is specified as a diameter in decimal degrees. The aperture parameter is only used for spectral extraction; a spectral extraction SSA service **must** support this parameter.

If no aperture is specified by the client the service should supply a default value appropriate to the data in question, for example, a circular aperture large enough to capture 98% of the signal from a point source in the aperture, knowing the spatial resolution of the data in the desired spectral band. The size of the aperture used to generate the simulated data should be returned in the description of the data in the query response table. The service should not normally use SIZE as or the default value of the aperture for spectral extraction, as this will generally represent an upper limit on the maximum separation of the target position and observed position when the SSA query is used to search for data potentially matching a given target object, not knowing whether a given service is searching pre-existing spectral collections or computing extracted spectra.

Note that SSA makes it possible to find data for a specific point source object on the sky merely by specifying the estimated object position. For catalog spectra, SIZE defaults to whatever is appropriate for a possible match for an object in the catalog. For extracted spectra, the measurement aperture should be a value judged to be appropriate for the spatial resolution of the data.

If these simple heuristics are not adequate, the client data analysis application should explicitly specify the diameter of the synthetic aperture to be used.

### 4.1.2.2  SPECRP

The minimum spectral resolution, specified as the spectral resolving power $\lambda/d\lambda$ in dimensionless units.

### 4.1.2.3  SPATRES

The minimum spatial resolution (corresponding to the PSF of the observed signal) specified in decimal degrees.

### 4.1.2.4  TIMERES

The minimum time resolution, specified in seconds. For a typical spectrum the time resolution corresponds to the bounds of the time coverage of the exposure.

### 4.1.2.5  SNR

The minimum signal-to-noise ratio of a candidate dataset, for example specified as the ratio of the mean signal to the RMS noise of the background (see the SSA data model document for more detailed recommendations on how to compute the SNR).

### 4.1.2.6  REDSHIFT

A photometric (observed) redshift range specified as a single element open or closed range-list as defined in section 8.7.2. A negative redshift indicates a "blueshift", e.g., an object in the local neighborhood with a proper motion towards the Earth (a negative redshift is not proper terminology but this is thought to be simpler than other alternatives such as defining new terminology or adding additional parameters). An open range may be used to specify a minimum or maximum value. The optical redshift convention should be used ($d\lambda/\lambda$).

> Example: `1.2/3`

### 4.1.2.7  VARAMPL

The acceptable range of variability amplitude, specified as a single element open or closed range-list, with values in the range 0.0 to 1.0.

### 4.1.2.8  TARGETNAME

The target name, suitable for input to a name resolver. In general it may be preferable to perform target name resolution on the client-side, using POS to drive the query performed by the service, so that any service can respond to the query. The main reason that TARGETNAME is included here is to make it possible to find spectra of objects that do not have a known position, for example, spectra of solar system planets or asteroids. For a service which can supply spectra for moving objects, TARGETNAME is a required parameter; for other SSA services it is not normally required, but can be provided as an optional capability. If both TARGETNAME and POS are specified, both must satisfy the query for a candidate object to be matched.

### 4.1.2.9  TARGETCLASS

A comma delimited list of strings denoting the types of astronomical objects to be searched for. At the moment there is no standard classification for astronomical objects but it is suggested to use the "condensed" names from the list at http://simbad.u-strasbg.fr/guide/chF.htx.

> Examples: `star`, `galaxy`, `pulsar`, `PN`, `AGN`, `QSO`, `GRB`

### 4.1.2.10 FLUXCALIB

Specifies the minimum level of flux calibration for acceptable data. Possible values are "`absolute`", "`relative`", "`normalized`", and "`any`" (the default). If "relative" is specified, spectra which have an absolute flux calibration will be found as well. "Normalized" refers to spectra which have been normalized by dividing by a reference spectrum (including continuum normalization).

### 4.1.2.11 WAVECALIB

Specifies the minimum level of spectral coordinate calibration for acceptable data. Possible values are "`absolute`", "`relative`", and "`any`" (the default). If "relative" is specified, spectra which have an absolute spectral coordinate calibration will be found as well.

### 4.1.2.12 PUBDID

The IVOA publisher's dataset identifier, assigned by the publisher of a dataset. The same dataset published in different places may have a different PUBDID assigned by each publisher, however, unlike CREATORDID, where data creators may often not assign IVOA identifiers; it is guaranteed that a publisher can always assign a unique PUBDID when a dataset is published to the VO. ADS dataset identifiers are an example of a PUBDID, but in general any publisher may assign their own unique publisher dataset identifier. Publisher dataset identifiers may be determined by a prior query or some external means, such as another form of archive query.

> **Note:**
> A special case of a publisher's dataset identifier is the **ADS dataset identifier,** used to reference published IVOA datasets in journal articles.

### 4.1.2.13 CREATORDID

An IVOA dataset identifier, assigned at creation time by the creator of the parent data collection (survey project, observatory, etc.). Datasets can have a globally unique CreatorDID assigned prior to publication of the data to the VO, for example when the data is generated by a processing pipeline, or ingested into the master archive for the data collection. This is possible since the Creator entity for a data collection (e.g., an observatory or survey project) controls its own namespace, which can be registered as a globally unique Authority identifier. When a CreatorDID has been assigned this is the most universal way to refer to a dataset, as all replicated versions will share the same CreatorDID regardless of where they are published. Creator dataset identifiers may be determined by a prior query or by some other means, such as another form of archive query.

Example: `ivo://nrao.edu/vla#1998s2/4992a`

### 4.1.2.14 COLLECTION

The IVOA identifier or "shortName" of a data collection as defined by the service, for example `SDSS-DR2`, or `NRAO-VLA`. By data collection we refer to an organized, uniform collection of datasets from a single source, for example a single data release from a survey, or an instrumental data collection from an observatory. Unless an IVOA identifier is input, the service should treat the search term as a case insensitive, minimum match string. For instance, "dss" would match either `dss1` or `ESO-DSS2`. Allowable data collection references are specified in the service capabilities.

### 4.1.2.15 TOP

TOP limits the number of returned records in the query response table to the specified number of top ranking ones.  Records are ranked according to a "score" heuristic (Dolensky 2006).  The details of the actual heuristic used are up to the service, but the general idea is that the better a candidate dataset matches the query, the higher the score it receives. Metrics such as distance from the specified position, or the degree of overlap with a specified bandpass or time interval, determine the score.  If two datasets would otherwise have the same score, the service may use other unspecified dataset characteristics, such as some intrinsic data quality metric, to further rank candidate datasets.  If the service implements a ranking heuristic the query response table should normally be returned sorted in order of decreasing score.  TOP can also be used by the client to limit the size of the query response table in cases where the query might find a very large number of candidate objects.

### 4.1.2.16  MAXREC

The maximum number of records to be returned.  This may be used to increase the built-in default limit set in the service, up to some maximum service-specified default (this is provided in an attempt to permit larger queries without having to page through the query response, which requires saving state on the server).  A service should typically have a fairly small default MAXREC, provided to improve the query response time, and a large upper limit on MAXREC, provided to enable large queries.

### 4.1.2.17 MTIME

Find only datasets modified, created, or deleted in the given range of dates, specified as a single element in range-list format, as an open or closed range, with the dates specified in ISO 8601 format.  Note this is not the same thing as TIME, which refers to time of observation.  MTIME may be used to periodically query services for new or updated data. Deleted datasets are indicated by a non-null deletion date in the Dataset.Deleted field of the query response.  Services which support MTIME should also support Dataset.Deleted (see also 4.2.5.4).

### 4.1.2.18 COMPRESS

If this flag is present, datasets returned via the *getData* method **may** optionally be returned to the client in compressed form.  Valid values are "true" and "false"; if the COMPRESS parameter is included in the query without a value, "true" is assumed.  By default compressed data is not permitted.

Compression is performed by applying a whole-file compression algorithm such as *gzip*, and updating the HTTP content type of the returned document accordingly.  Dataset-level compression is distinguished from protocol-level compression, which is performed at the level of the HTTP protocol, on the entire data stream, and is transparent to the client. Support for compression as a data format option (e.g., FITS HCOMPRESS) is not yet defined or supported at this time.

### 4.1.2.19 RUNID

The RUNID is an opaque string used to associate multiple service invocations in service logs, e.g., to identify them as all belonging to the same job or application. RUNID is not used by SSA in any way, except in cases where SSA may call another VO service, in which case the RUNID parameter **should** be passed on to the called service. The purpose of RUNID is to allow the job run ID to be logged, and in particular, if a job involves multiple requests to multiple services, to allow all just requests to be associated by having a common RUNID.

## 4.1.3 Service-Defined Parameters

The service **may** support additional service-defined parameters. Parameter names must not match any of the reserved parameter names defined herein, independent of case.

Any service defined parameters should be defined in the metadata query response (6.1). Appendix A presents an example of this, where service defined parameters are used to dynamically generate spectra based upon a theoretical model.

## *4.2 Query Response*

The output returned by a query is an XML document compliant with VOTable V1.1 or greater (VOTable 2004) and should be returned with a base MIME-type of `text/xml` to enable simple display of query results in browsers using direct rendering of the XML, or an optional style sheet. Parameterization may be used to further refine the MIME-type, for example "`text/xml;content=x-votable`" may be used to indicate that the content of the XML document returned is a VOTable.

> **Note:**
> The `FORMAT` parameter has no influence on the query response. `FORMAT` applies only to the returned datasets, not to the query response. The query response is always returned as a VOTable.

The VOTable **must** contain a RESOURCE element, identified with the tag `type = "results"`, containing a single TABLE element with the results of the query. Additional RESOURCE elements may be present, but the usage of any such elements is not defined here.

The `RESOURCE` element **must** contain an INFO with `name="QUERY_STATUS"`. Its value attribute should be set to ″OK″ if the query executed successfully, regardless of whether any matching data were found. All other possible values for the value attribute are described in section 8.10.

> **Examples:**
> ```
> <INFO name="QUERY_STATUS" value="OK"/>
> <INFO name="QUERY_STATUS" value="OK">Successful Search</INFO>
> ```

Another INFO with name="SERVICE_PROTOCOL" should contain the protocol version number in its value attribute and the name of the service protocol as the fixed string "SSAP" (see version negotiation 8.2.4).

**Example:**
```
<INFO name="SERVICE_PROTOCOL" value="1.02">SSAP</INFO>
```

Additional INFOs may be provided, e.g., to echo the input parameters back to the client in the query response (a useful feature for debugging or to self-document the query response), however this is not required.

In the query response table each row represents a different physical or virtual dataset which is potentially available to the client. The VOTable GROUP construct is used to associate related groups of fields. Table FIELDs describe the attributes of each dataset; if all datasets share the same value for an attribute it can be represented as a PARAM.

**Hint:**
Put constant values in PARAM elements instead of repeating them in each table row.

### 4.2.1 Query Response Metadata

Names of fields and parameters are left to the service provider. UTYPEs of standard fields are required for identification of interface elements and **must** be given and **must** comply with the SSA protocol (this document) and the Spectrum data model (McDowell, Tody, et. al. 2007). UCDs **should** also be given when specified by the protocol (not all interface or data model elements have assigned UCDs), but are not used to identify interface or data model elements. Values for the UCDs of standard interface and data model elements, where defined, are given in this specification and in the Spectrum data model document.

**Note:**
**UTYPE** values **must** be provided to identify interface or data model elements.
**UCD** values for standard data model elements **should** be provided as well.
Omit the leading "spectrum." In the UTYPE for Spectrum data model attributes.

The SSA query response consists of a number of fields, identified by UTYPE, grouped into component data models of the form "*<component-name>*'.'*<field-name>*". Some components of the query response are defined directly by the SSA protocol (this document), while others are taken directly from the Spectrum data model. Unless otherwise specified, the leading "spectrum." in the UTYPE values specified in the Spectrum data model is omitted in the SSA query response since this metadata is not specific to spectra and we use the same metadata for other types of data objects. Hence most of the query response metadata consists of generic component data models. For example, if the Spectrum data model specifies Spectrum.Target.Name this appears in the SSA query response as Target.Name. Applications can refer to Target.Name regardless of whether the data to be accessed is a Spectrum or some other data object such as an Image.

In the following, query response parameters which are mandatory, recommended, or optional are indicated as such in the tables or specified more precisely in the text.

Additional attributes from the Spectrum data model not shown here may appear in the query response table. The SSA query response does not include any actual data values, and elements of the Spectrum data model used to represent data values are not included here (the client needs to download the full dataset to get the data).

When a generic data model is applied in a specific context, the requirements for what is required, what is optional, and flexibility in what is permitted will vary depending upon how the data model is being used. Hence when data model attributes are indicated as **mandatory or recommended** in this document, this overrides any similar requirements specified in the Spectrum data model document. The SSA query response is also more restrictive than the underlying model; in particular the **allowable units** are more restrictive than what is permitted in the model, in order to be more consistent with other elements of SSA, and to provide more uniformity to make multiband data discovery by the client easier. Hence within SSA, characterization restricts the allowable units for spatial coordinates to decimal degrees, for spectral coordinates to wavelength in meters, and for time measures to seconds, except where MJD is used (all represented in floating point).

It is difficult to specify every detail of every metadata element in this document without burdening the text with too much detail; furthermore, many optional metadata values are omitted from the summary tables shown here. Full details are given in the Spectrum data model document, and in a convenient summary form in a spreadsheet which lists all metadata elements with full details for each. All this information can be found on the SSA TWiki page at http://www.ivoa.net/twiki/bin/view/IVOA/SsaInterface.

Query metadata may be mapped to VOTable fields in any order, so long as fields which are part of the GROUP construct (all the component data models are GROUP elements) appear in consecutive table fields.

## 4.2.2 Types of Metadata

Metadata in the query response is grouped into a number of component data models as summarized in the table below, and explained in more detail in the sections which follow.

| Service Metadata | |
|---|---|
| Query | Describes the query itself |
| Association | Logical associations |
| Access | Dataset access-related metadata |
| | |
| **Data Model Metadata** | |
| Dataset | General dataset metadata |
| DataID | Dataset identification (creation) |
| Curation | Publisher metadata |
| Target | Observed target, if any |
| Derived | Derived quantities |
| CoordSys | Coordinate system frames |
| Char | Dataset characterization |
| | |
| **Characterization Metadata** | |
| Char.FluxAxis | Observable, normally a flux measurement |

| Char.SpectralAxis | Spectral measurement axis, e.g., wavelength |
|---|---|
| Char.TimeAxis | Temporal measurement axis |
| Char.SpatialAxis | Spatial measurement axis |

**Service metadata** is specific to the functioning of the service itself, for example to step through large queries or retrieve selected datasets. **Data model metadata** describes each dataset, and is common between the SSA protocol and the Spectrum data model. **Characterization metadata** physically characterizes the dataset in terms of the spatial, spectral, and temporal measurement axes and the observable. Characterization is part of the data model but is broken out separately in the table above to show the major elements of the characterization model. Most of the metadata returned by SSA is generic dataset metadata, which means it is not actually specific to spectra and may be used in other DAL interfaces to describe other types of dataset, for example an image or catalog. For data model metadata, please refer to the Spectrum data model for details such as the UCD and units, unless specified otherwise in this document.

Each of these types of query response metadata is discussed in more detail in the sections which follow.

### 4.2.3 Query Metadata
Query metadata describes the query itself.

| UTYPE | Description | Req |
|---|---|---|
| Query.Score | Degree of match to query params | REC |

#### 4.2.3.1 Query.Score
A record with a higher score more closely matches the query parameters. The score is expressed as a floating point number with an arbitrary scale (different queries may return results with different scale factors and cannot be inter-compared). If scoring is used, the query response table should be returned sorted in order of decreasing values of score, with the top-scoring items at the top of the list. The details of the heuristic used to compute the score are left to the service. See the discussion of the TOP parameter in section 4.1.2.15.

### 4.2.4 Association Metadata
Association metadata is used to describe logical associations relating datasets described in the query response, as described in section 2.10. Logical associations between query response records may refer to the data access operation itself, e.g., where the same data object is available in multiple output formats, or to logical associations relating the physical data, e.g., where multiple primary datasets are part of the same observation. The same dataset may belong to multiple associations.

| UTYPE | Description | Req |
|---|---|---|
| Association.Type | Type of association | OPT |
| Association.ID | Unique ID identifying the association instance | OPT |
| Association.Key | Unique key different for each element of association | OPT |

Each such association is described by a separate instance of the Association model, with a defined Association Type, ID, and Key. In many cases the Association Type and Key can be represented as fixed PARAMs, leaving only Association.ID to be represented as a FIELD in each table row.

In general, specification of the allowable Assocation types is beyond the scope of this specification. The semantic details of Associations are intended to be defined either at a lower level, for a specific data collection or service, or at a higher level, e.g., to describe complex data associations. An exception is the MultiFormat association described in the next section.

### 4.2.4.1 MultiFormat Association

A pre-defined case is the *MultiFormat* association, where several records refer to the same dataset which is available in several different output data formats. In this case Association.Type should be set to "`MultiFormat`", Association.ID can be anything (an example is given in Appendix B where the values are of the form "MultiFormat.*<counter>*"), and Association.Key should be set to "`@Access.Format`" to indicate that the key which differentiates the elements of the association is the value of the Access.Format field of the record. If several query response records are of this type the association **should** be specified to indicate the association. In all other cases (currently undefined by the protocol) the association **may** be specified.

### 4.2.4.2 Association.Type

A service-defined type used to indicate what type of association is being referred to. The value should be unique within the scope of the query response. There can be many types of logical associations. Associations provide a means of describing complex data aggregations relating multiple datasets (spectra in the case of SSA). Association is a type of extension mechanism, and the definition of associations is beyond the scope of SSA; SSA merely provides the means to define and manipulate associations. Examples of possible associations might be an Echelle observation consisting of 100 orders, each of which appears in the query response as an individual 1-D spectrum, or a group of query response records which all refer to the same dataset but differ only in the output format.

Since the association type may be shared by many table records, it may be best specified as a PARAM in the output VOTable, using an ID-REF to link it to the association it refers to. An association type **should** be provided for each association in the table.

### 4.2.4.3 Association.ID

The association ID is a string, unique within the scope of a given VOTable, identifying one instance of a given association. All members of the association instance share the same Association.ID. The association ID **must** be provided for any association. The content of the string is up to the service. Multiple association IDs may be provided for a single record if a record belongs to more than one association. Note that Association.ID is unrelated to the VOTable ID, which is used to uniquely identify the elements of a VOTable.

Extension metadata may optionally be provided to describe an association in more detail. Extension metadata appears in the output VOTable as optional additional RESOURCE elements (see section 4.2.7). The ID-REF mechanism may be used to link such an extension record to the association in the main table. The contents of an association metadata extension record are externally defined and beyond the scope of SSA.

### 4.2.4.4  Association.Key

The association key **should** be provided to identify what is "different" for each member of an association. The value is a string and may be either an arbitrary value defined by the association, or a reference to one or more table fields which form the association key. If a table field is referenced the '`@`' character should be prefixed to the VOTable ID of the referenced FIELD to indicate the indirection (e.g., "`@Format`"), otherwise the literal string is used as the key. A key may contain multiple elements delimited by commas.

## 4.2.5 Access Metadata

Access metadata is required to tell a client how to access the datasets described in the SSA query response.

| UTYPE | Description | Req |
|---|---|---|
| **Access.Reference** | URI (URL) used to access the dataset | MAN |
| **Access.Format** | MIME type of dataset | MAN |
| Access.Size | Estimated (not actual) dataset size | REC |

### 4.2.5.1  Access Reference

The access reference is a URI (typically a URL) which can be used to synchronously retrieve the specific dataset described in a row of the query table response. If the dataset pointed to by the access reference does not exist at query time, it will be computed on the fly when accessed.

Since the datasets supported by SSA are typically small (compared to images), SSA does not currently support data staging and asynchronous data access. Support for this may be added in a future version, e.g., to support generation of simulated or synthetic data.

Since the access reference is a URL, it is convenient to be able to input the access reference directly in a Web browser or other standard Web tool to access the referenced dataset. For this reason the access reference string should be URL-encoded if it contains any reserved URL metacharacters (the "#" character used in dataset identifiers is particularly nasty). See also section 8.3.2. The CDATA construct used in earlier data access interfaces (SIAP V1.0) does not serve the same purpose and should not be used; use URL encoding instead.

### 4.2.5.2  Output Format

The file format of a candidate dataset is specified by its MIME type. Both uncompressed and compressed data can be indicated in this fashion.

The file format says nothing about the data model used by whatever data object is stored in the file; this is specified by the Dataset.DataModel attribute discussed in section 4.2.5.4.

A single data object may be available in multiple file formats. In such a case an association **should** be defined to indicate that the entries all refer to the same data object.

### 4.2.5.3  Dataset Size Estimate

The approximate estimated size of the dataset, specified in kilobytes, **should** be given to help the client estimate download times and storage requirements when generating execution plans. Only an approximate, order of magnitude value is required (a value rounded up to the nearest hundred KB would be sufficient). In the VO dataset sizes can vary by many orders of magnitude hence it is important to know this information to optimize execution plans before attempting to download data or request computation. It is preferable to return an order of magnitude estimate of the dataset size, than no value at all. A precise value is *not* required.

## Data Model Metadata

The following metadata components are in common with the Spectrum data model.

### 4.2.5.4  General Dataset Metadata

General dataset metadata describes the overall dataset.

| UTYPE | Description | Req | Default |
|---|---|---|---|
| **Dataset.DataModel** | Datamodel name and version | MAN | Spectrum-1.0 |
| Dataset.Type | Type of dataset | OPT | Spectrum |
| **Dataset.Length** | Number of points in spectrum | MAN | |
| Dataset.Deleted | Set to deletion time, if dataset is deleted | OPT | |
| Dataset.TimeSI | SI factor and dimensions | OPT | |
| Dataset.SpectralSI | SI factor and dimensions | OPT | |
| Dataset.FluxSI | SI factor and dimensions | OPT | |
| Dataset.SpectralAxis | SpectralAxis column name (native data) | OPT | |
| Dataset.FluxAxis | FluxAxis column name (native data) | OPT | |

Dataset.DataModel is a string identifying the data model type and version used in the described dataset. For SSA-compliant data this should be a value such as "`Spectrum-1.0`", as specified in the Spectrum data model document for the version of the data model being used. For pass-through of native project data some other value should be used which identifies the specific project data model used, e.g., "`HST-STIS-1.0`".

For the current SSA interface, Dataset.Type  is always "`Spectrum`" and can normally be omitted. Dataset.Length is mandatory and specifies the "length" of the spectrum, i.e., the number of data points or samples.   Dataset.Deleted is used with the MTIME query parameter to inform the client that a previously existing dataset has been deleted; if a service supports MTIME it should also support Dataset.Deleted. The value is the ISO 8601 date (as in MTIME) at which the dataset was deleted, or null for a normal non-deleted dataset. Dataset.Deleted should be returned in a query only if MTIME is used in the query, and the deletion date matches the interval of time specified by MTIME. Otherwise deleted datasets should never be visible in a query. A service may permanently delete dataset deletion history after a period of time (currently unspecified) long enough to permit clients to discover deleted datasets.

The SI parameters provide a simplified approach to defining the units of the spectral coordinate and flux density, e.g., for overplotting spectra from different sources. Each SI parameter shall conform to the description in the Spectral Data Model standard, section 3.2 "Units" (McDowell/Tody et al. 2007). Dataset.SpectralAxis and Dataset.FluxAxis are used to identify the spectral and flux axes in native format data, where spectra are returned in a project-specific tabular data format such as FITS binary table; the values specify the names of the table columns used. While the SI and Axis parameters are optional if the full Spectrum data model is used, they **should** be provided for native data to make it possible for a client to interpret such data in a basic fashion without having to understand the details of each project-specific native data format.

### 4.2.5.5 Dataset Identification Metadata

Dataset identification metadata is used to describe the fundamental identify of a dataset, including where it came from and how it was created.

| UTYPE | Description | Req | Default |
|---|---|---|---|
| **DataID.Title** | Dataset title | MAN | |
| DataID.Creator | Creator name (string) | REC | |
| DataID.Collection | IVOA Identifier of collection (string) | REC | |
| DataID.DatasetID | IVOA Dataset ID | OPT | |
| DataID.CreatorDID | Creator assigned dataset identifier | REC | |
| DataID.Date | Data processing/creation date | OPT | |
| DataID.Version | Version of creator-produced dataset | OPT | |
| DataID.Instrument | Instrument name | OPT | |
| DataID.Bandpass | Bandpass name, e.g., filter | OPT | |
| DataID.DataSource | Original source of data | REC | survey |
| DataID.CreationType | Dataset creation type | REC | archival |

Dataset.Title is a short, human-readable description of a dataset, and should be less than one line of text. Information such as the instrument or survey name, filter, target name, etc., is typically included in a condensed form. The exact contents of Dataset.Title are up to the data provider. Dataset.Creator identifies the entity which created the dataset, and should be a short string consistent with the RSM specification, e.g., `SDSS`. Dataset.Collection is the registered IVOA identifer of the data collection to which the dataset belongs, e.g., `ivo://sdss/dr5/spec`.

The CreatorDID is the IVOA dataset identifier (if any) assigned by the entity which created the dataset *content*, typically (but not always) an observatory or survey project. If the dataset referred to is virtual data, CreatorDID refers to the parent dataset from which the virtual data will be created (see 2.5.2 for further details). If a CreatorDID has been assigned to a dataset it **should** be provided, otherwise it should be omitted. DataID.Date, specified in ISO time format, specifies the date when the dataset was created or last modified by the DataID.Creator entity. If a dataset is modified or replaced without changing its CreatorDID, DataID.Date and DataID.Version should be updated accordingly.

DataID.Instrument is a short string identifying the instrument used to create the data (instrument may be an actual telescope instrument or something else, e.g., a program in the case of theory data). DataID.Bandpass is a short string specifying the bandpass name if any, e.g., a filter name or an instrumental bandpass such as I, J, K, Q, HI, and so forth. Values specified with DataID.Bandpass may be used as input to the BAND parameter (4.1.1.3) to refine a query (if this feature is supported by the service).

DataID.DataSource and DataID.CreationType describe the original source of the data, and how the dataset returned by the service was or will be created, as defined in section 2.5.

### 4.2.5.6 Curation Metadata

Curation metadata describes who curates the dataset and how it is published to the VO.

| UTYPE | Description | Req | Default |
|---|---|---|---|
| **Curation.Publisher** | Publisher | MAN | |
| Curation.Reference | URL or Bibcode for documentation | REC | |
| Curation.PublisherDID | Publisher's ID for the dataset | REC | |
| Curation.Date | Date curated dataset last modified | OPT | |
| Curation.Version | Version of curated dataset | OPT | |
| Curation.Rights | Restrictions: public, proprietary, etc | OPT | public |

Curation.Publisher is a short string identifying the publisher of the data, e.g., a data archive or data center, or an indexing service such as the ADS. Curation.PublisherDID is the IVOA dataset identifier (URI) assigned by the publisher to identify the dataset within its holdings. Curation.Reference is a forward link to publications which reference the dataset; multiple instances are permitted. Curation.Date and Curation.Version refer to the dataset *as curated by the publisher*, hence can differ from the same values given in DataID, which refer to the *content* of the dataset as generated by the dataset Creator. Curation.Rights specifies whether the dataset is "public" or "proprietary". Proprietary data requires authentication and authorization by the data provider to access, and once downloaded should be protected from subsequent access on the client side.

> **Note:**
> If the same dataset is replicated at several locations with multiple publishers, it is possible to set up an association group to indicate this fact.

### 4.2.5.7 Astronomical Target Metadata

Target metadata describes the astronomical target observed, if any.

| UTYPE | Description | Req | Default |
|---|---|---|---|
| Target.Name | Target name | OPT | |
| Target.Class | Target or object class | OPT | |
| Target.Redshift | Target redshift | OPT | |
| Target.VarAmpl | Target variability amplitude, typical | OPT | |
| Derived.SNR | Signal-to-noise for spectrum | OPT | |

Target.Name is a short string identifying the observed astronomical object, suitable for input to a name resolver. Target.Class is the object class if known, e.g., Star, Galaxy, AGN, QSO, and so forth (see section 4.1.2.9). Target.Redshift, Target.VarAmpl, and Derived.SNR are as defined in the data model. Either standard target values, or derived quantities, may be used in the query response.

### 4.2.5.8 Coordinate System Metadata

Coordinate system metadata describes the coordinate system reference frames used in the SSA query response.

| UTYPE | Description | Req | Default |
|---|---|---|---|
| CoordSys.SpaceFrame.Name | Spatial coordinate frame | REC | ICRS |
| CoordSys.SpaceFrame.Equinox | Equinox | OPT | 2000.0 |
| CoordSys.TimeFrame.Name | Timescale | OPT | TT |
| CoordSys.TimeFrame.Zero | Zero point of timescale in MJD | OPT | 0.0 |

These reference frames apply to all spatial (sky), spectral, and time coordinates used in the SSA query response (including Characterization) unless otherwise specified. Note that spatial coordinates are not limited to the celestial sphere; any spatial coordinate frame specified in the data model may be specified, including solar and planetary coordinate systems, although the default is ICRS.

### 4.2.5.9 Dataset Characterization Axis Metadata

The Characterization axis metadata specifies the type of physical quantity on each physical measurement axis as well as the observable.

| UTYPE | Description | Req | Default |
|---|---|---|---|
| Char.FluxAxis.Ucd | ucd for flux | REC | |
| Char.SpectralAxis.Ucd | ucd for spectral coord | REC | |

Values are specified as UCDs, as defined in the data model. For example, to specify that the flux axis is flux density per unit wavelength, the value "`phot.fluDens;em.wl`" would be given.

### 4.2.5.10 Characterization Coverage Metadata

The Coverage component of the Characterization data model (Char) describes the coverage of the dataset in each of the three primary measurement axes.

| UTYPE | Description | Req | Default |
|---|---|---|---|
| **Char.SpatialAxis.Coverage.Location.Value** | Observed position, e.g., RA DEC | MAN | |
| **Char.SpatialAxis.Coverage.Bounds.Extent** | Aperture angular diameter, deg | MAN | |
| Char.SpatialAxis.Coverage.Support.Area | Aperture region | OPT | |
| Char.SpatialAxis.SamplingPrecision.FillFactor | Sampling filling factor | OPT | 1.0 |
| **Char.TimeAxis.Coverage.Location.Value** | Midpoint of exposure (MJD) | MAN | |
| Char.TimeAxis.Coverage.Bounds.Extent | Total elapsed exposure time | REC | |
| Char.TimeAxis.Coverage.Bounds.Start | Start time | OPT | |
| Char.TimeAxis.Coverage.Bounds.Stop | Stop time | OPT | |
| Char.TimeAxis.Coverage.Support.Extent | Effective exposure time | OPT | |

| Char.TimeAxis.SamplingPrecision.FillFactor | Sampling filling factor | OPT | 1.0 |
|---|---|---|---|
| **Char.SpectralAxis.Coverage.Location.Value** | Midpoint of Spectral coord range | MAN | |
| **Char.SpectralAxis.Coverage.Bounds.Extent** | Width of spectrum in meters | MAN | |
| Char.SpectralAxis.Coverage.Bounds.Start | Start in spectral coordinate | REC | |
| Char.SpectralAxis.Coverage.Bounds.Stop | Stop in spectral coordinate | REC | |
| Char.SpectralAxis.SamplingPrecision.FillFactor | Sampling filling factor | OPT | 1.0 |

Within Char, Coverage specifies the *location* (central or characteristic value), *bounds* (measurement limits), *support* (region covered within the bounds), and *filling factor* (fraction of total area covered) for each measurement axis. The coordinate system reference frames specified in Coordsys apply here. Spatial coordinates are specified in units of decimal degrees, spectral coordinates in units of meters, and time coordinates in units of days.

### 4.2.5.11 Characterization Accuracy and Error Metadata

The `Accuracy` component of Characterization specifies the sampling, resolution, and error estimates for the dataset.

| UTYPE | Description | Req | Default |
|---|---|---|---|
| Char.FluxAxis.Accuracy.StatError | Statistical error | OPT | |
| Char.FluxAxis.Accuracy.SysError | Systematic error | OPT | |
| Char.FluxAxis. Calibration | Type of flux calibration | REC | calibrated |
| Char.SpectralAxis.Accuracy.BinSize | Wavelength bin size | OPT | |
| Char.SpectralAxis.Accuracy.StatError | Spectral coord measurement error | OPT | |
| Char.SpectralAxis.Accuracy.SysError | Spectral coord measurement error | OPT | |
| Char.SpectralAxis. Calibration | Type of coord calibration | REC | calibrated |
| Char.SpectralAxis.Resolution | Spectral resolution FWHM | REC | *BinSize* |
| Char.TimeAxis.Accuracy.BinSize | Time bin size | OPT | |
| Char.TimeAxis.Accuracy.StatError | Time coord statistical error | OPT | |
| Char.TimeAxis.Accuracy.SysError | Time coord systematic error | OPT | |
| Char.TimeAxis. Calibration | Type of coord calibration | OPT | calibrated |
| Char.TimeAxis.Resolution | Temporal resolution FWHM | OPT | *BinSize* |
| Char.SpatialAxis.Accuracy.StatError | Astrometric statistical error | REC | |
| Char.SpatialAxis.Accuracy.SysError | Systematic error | OPT | |
| Char.SpatialAxis. Calibration | Type of coord calibration | REC | calibrated |
| Char.SpatialAxis.Resolution | Spatial resolution of data | REC | |

Both overall statistical and systematic error estimates may be specified. The calibration status of all three primary measurement axes as well as the observable **should** be given, otherwise "calibrated" is assumed. The spatial and spectral resolution **should** be specified. Note that, for consistency within Char, the spectral resolution is specified here in spectral coordinate units (FWHM in meters), unlike the SPECRP query parameter, which is specified as λ/dλ.

### 4.2.6 Additional Service-Defined Metadata

A given service **may** return additional query response metadata not defined by the SSA protocol. This additional metadata may take the form of additional table columns, or additional RESOURCE elements in the query response VOTable.

Service-defined output metadata **should** use service-defined UTYPEs and UCDs as long as they do not clash - and can be easily distinguished - from mandatory and reserved SSA output columns.

### 4.2.7 Metadata Extension Mechanism

The metadata extension mechanism allows a data provider to add additional custom metadata to the query response to describe collection-specific details of the data. Extension metadata appears in the query response as additional RESOURCE elements in the VOTable. The format and contents of these RESOURCE elements is up to the data provider. The ID-ref mechanism of VOTable is used to link extension elements to associated fields of the main query response VOTable.

The extension RESOURCE elements can contain PARAMs, TABLEs, or nested RESOURCEs. The ID-ref mechanism simply allows associating FIELDs from the main table to RESOURCEs, TABLEs and GROUPs in the extension RESOURCEs. It is actually the core of an indexing mechanism where the values of the referring FIELD are used as a key to associate a specific query response field to some additional information. In case the referred element is a TABLE in the extension RESOURCE, this TABLE must contain a FIELD identical to the referring FIELD and the indexing mechanism will provide a classical "a la RDBMS" jointure. In case the referred element is a RESOURCE of the extension, the key value is assumed to be the ID of a nested element (PARAM, RESOURCE, or TABLE) which is associated to the FIELD of the main table. It is also possible to provide backward linking by referring to FIELDs in the main section from elements in the extension.

As for any VOTable, the client software is guided for the usage of these extensions by the UTYPEs of the main query response and extension elements. UTYPE identifies the exact meaning of the element in a specific data model. In the context of the DAL metadata extension mechanism, UTYPE gives the meaning of the association mechanism described above, and of all the extension elements. Current available IVOA data model and UTYPEs are defined for resource metadata, VOEvent, STC, Characterization and the Spectrum data model. It is possible to use additional adhoc UTYPEs by agreement between data providers and client developers. These adhoc UTYPEs can be described as conventions in IVOA notes, or may be replaced eventually by new IVOA standard UTYPEs when these become available.

## 5 GetData (reserved)

The current SSA protocol does not include an explicit specification for a *getData* operation; an access reference URL is used instead, to provide maximum flexibility in how the *queryData* operation refers datasets back to the service for access. A more explicit getData operation remains an option in the future for accessing referenced datasets. This would still allow a URL to be used as at present, but could provide finer control over access-related options such as the output data format, or the ability to refer to a virtual dataset externally by a publisher assigned dataset identifier.

# 6  Metadata Query

## 6.1 Metadata Request

A compliant service **must** support queries with `FORMAT=METADATA` used to query the service metadata; only metadata is returned by the service in this case. The `FORMAT=METADATA` query is implemented as a special case of the queryData operation, using the FORMAT parameter. When `FORMAT=METADATA` is given, all other queryData input parameters **should** be ignored (except REQUEST and VERSION, which are common to all operations and not specific to queryData). The response to this query provides two types of information about the service:

- supported input parameters (4.1)
- possible output columns (4.2)

Note that the SSA specification is designed so that a client does not need to know this information to make use of the service. It is most useful for communicating non-standard or custom input and output parameters. The metadata query can also be helpful to a registry for verification purposes.

> **Note:**
> A *getCapabilities* operation returning service metadata will be added in a future version which will obsolete the current `FORMAT=METADATA` mechanism. As an additional service operation, the addition should be backwards compatible with existing services.

## 6.2 Metadata Response

The structure of the VOTable returned by a metadata query (see sample in Appendix C) is similar to a normal queryData response. In particular it is also similar to SIA V1.0 (Tody/Plante 2004) except that it requires VOTable V1.1+ (VOTable 2004) which supports the UTYPE attribute to link it up with the spectral data model (McDowell/Tody et al. 2007).

Also the input parameters supported by the service **must** be listed as a PARAM element of the RESOURCE that normally contains the query response table, including required parameters (4.1.1), optional parameters (defined in 4.1.2) and non-standard parameters specific to the service (4.1.3).

The VOTable format mandates the presence of NAME, VALUE and DATATYPE attributes in PARAM elements. The NAME attribute **should** have the form `"INPUT:param_name"`, or `"OUTPUT:param_name"` where *param_name* is the parameter name as it appears in the query. For example, `name="INPUT:POS"` refers to the POS input parameter. The VALUE attribute **may** contain the default value that will be assumed if the parameter is not set in the query. Otherwise an empty `value=""` **must** be given because VOTable requires this attribute.

Each PARAM **should** have a UNIT attribute for the physical unit. Output parameters **should** also have a UTYPE attribute of the corresponding data model item where applicable. Implementors are encouraged to include, as children of the PARAMs,

DESCRIPTION elements to describe the parameter and (where appropriate) VALUEs elements to given allowed ranges or values.

Summary of PARAM attributes in the metadata response:

| PARAM Attribute | Remarks | Req | Sample/Template Value |
|---|---|---|---|
| name | name with prefix „INPUT\|OUTPUT" | MAN | INPUT\|OUTPUT:param_name |
| value | required by VOTable standard; may contain service default for input params; empty for output params | MAN | ALL |
| datatype | required by VOTable standard | MAN | double |
| unit | physical unit where applicable | REC | deg |
| utype | pointer to data model element | REC | ssa:Access.Reference |
| ucd, arraysize, precision, width, ... | copy from `queryData` response | OPT | meta.id |

> **Hint:**
>
> A quick way to build a metadata response document is to take an empty query result (without matching records) as a template and to add supported input parameters. Replace FIELD by PARAM elements. Patch the PARAM name attributes (INPUT/OUTPUT:param_name), add DESCRIPTION elements and arrange the order of PARAM elements according to the document structure in sections `4.1` and `4.2`.

# 7 Data Retrieval

The data retrieval request allows a client to retrieve a single spectrum given an access reference (ACREF) as returned by a prior queryData operation.

## 7.1 Access Reference URL

The access reference is a simple URL (IETF RFC 1738). In principle the URL may reference transports other than HTTP but at the present time this is not recommended. If the client issues a HTTP GET request using this URL, and the request is successful, the client will receive a document of the type given in query response column with the UTYPE Access.Format. Since a prior query to the service is required to obtain an ACREF, no requirements are placed on the form the ACREF takes; this is completely up to a given service implementation. This has the effect of hiding the details of the ACREF URL from the client, making it easy to layer an implementation of the GET web method on top of an existing data retrieval service, and making it easier to hide changes to the implementation of existing services.

## 7.2 Data Format

The response to a data retrieval request is a single Spectrum instance. Both the data mode of the returned spectrum, and the file data format, may vary, but must agree with what is specified in the query response. The data model of a spectrum will be either some version of the Spectrum data model, or an externally-defined data model such as the native project data model. The available file formats will in general depend upon the data model, but for Spectrum they include at least the following:

- `application/x-votable+xml`
- `application/fits`
- `application/xml`
- `text/csv` — comma separated values
- `image/jpeg` - graphics preview
- `text/html`

The graphics formats and text/html, if available, provide a directly viewable, rendered version of the spectrum. All the other formats return science data.

## 7.3 Data Compression

If the query parameter COMPRESS is present then the service **may** return a compressed dataset, using some standard compression technique such as *gzip*, in place of a normal dataset, without indicating this in the query response. Basically the client is indicating that it is prepared to receive either compressed or uncompressed datasets and does not care which is delivered (the service should pick whichever is more efficient). This should be distinguished from protocol-level compression, which is transparent to the client, and may occur at the level of the HTTP protocol if both client and server support HTTP protocol compression.

In case of an HTTP GET the keyword `Content-Encoding` informs the receiver about the encoding of the output data, and should have a value such as `gzip`. Note that the encoding is distinct from the MIME-type (`Content-Type`) of the returned data object.

## 7.4 Error Response

If possible, unsuccessful data retrieval should return a standard VOTable-format service error response, as outlined in section 8.10. Depending upon the nature of the error or how data retrieval is implemented, this may or may not be possible, and a HTTP error may result instead. The client should be prepared to handle either form of error. In particular, if the operation is successful at the HTTP level, the client must check for a VOTable error response to be sure that an error has not occurred.

# 8  Basic Service Elements

## 8.1 Introduction

This clause specifies aspects of SSA service behaviour that are independent of particular operations or are common to several operations.

## *8.2 Version numbering and negotiation*

### 8.2.1 Version number form and value

The SSA protocol defines a protocol version number. The version number applies to all aspects of the protocol as defined in this document, including any associated XML schema and the request encodings.  While the SSA protocol and the associated SSA (Spectrum) data model are separately versioned, they are not independent, and a given version of SSA assumes a compatible version of the Spectrum data model.  In other words,  knowing the SSA protocol version the client can assume a compatible version of the data model.

Version numbers follow IVOA document conventions and contains two non-negative integers, separated by decimal points, in the form *"x.y"*, for example, "1.0", or "1.13".  This is actually a *three* level version number encoded as two digits, e.g., "1.23" is logically the same as "1.2.3".  One result of this syntax is that second level version numbers cannot be greater than 9, for example "1.9" is a *higher* version number than "1.10" (logically "1.9.0 vs. "1.1.0").  Hence IVOA version numbers cannot be numerically compared without first being parsed.

### 8.2.2 Version number changes

The protocol version number shall be changed with each revision of this document. The number shall increase monotonically and shall comprise no more than two integers separated by decimal points, with the first integer being the most significant.  There may be gaps in the numerical sequence. Some numbers may denote draft versions. Servers and their clients need not support all defined versions, but shall obey the negotiation rules below.

A version number change at the first level (1.0 – 2.0) indicates a major change.  A version number change at the second level indicates a minor change which is not necessarily backwards compatible.  A version number change at the third level is considered backwards compatible, and should not affect the pre-existing functionality of the interface.

### 8.2.3 Appearance in requests and in service metadata

The version number may appear in at least three places: in the service metadata, the parameter list of client requests to a server and in the query response. The version number used in a client's request of a particular server shall be equal to a version number which that server has declared it supports (except during negotiation, as described below). A server may support several versions, whose values clients may discover according to the negotiation rules.

### 8.2.4 Version number negotiation

If a SSA client does not specify the version number in a request, the server assumes the highest *standard* version supported by the service, and no explicit version checking takes place.   If the client specifies an explicit version number, and this does not match a version available from the service at level two, the service returns a version number mismatch error. The client can determine what versions of the protocol the service supports by a prior call to getCapabilities (once this is specified) or via a registry query.

## *8.3 General HTTP request rules*

### 8.3.1 Introduction

This document defines the implementation of the SSA service on a distributed computing platform (DCP) comprising Internet hosts that support the Hypertext Transfer Protocol (HTTP) (see IETF RFC 2616). Thus, the Online Resource of each operation supported by a server is an HTTP Uniform Resource Locator (URL). The URL may be different for each operation, or the same, at the discretion of the service provider. Each URL shall conform to the description in IETF RFC 2616 (section 3.2 "HTTP URL") but is otherwise implementation-dependent; only the query portion comprising the service request itself is defined by this document.

While the SSA protocol currently only supports HTTP as the DCP for general parameterized operations, data access references are more general and may use other internet protocols, e.g., FTP, or potentially grid protocols.

HTTP supports two request methods: GET and POST. One or both of these methods may be offered by a server, and the use of the Online Resource URL differs in each case. Support for the GET method is mandatory; support for the POST method is optional.

### 8.3.2 Reserved characters in HTTP GET URLs

The URL specification (IETF RFC 2396) reserves particular characters as significant and requires that these be escaped when they might conflict with their defined usage. This document explicitly reserves several of those characters for use in the query portion of SSA requests. When the characters "?", "&", "=", "," (comma), "/", and ";" appear in one of the roles defined in Table 1, they shall appear literally in the URL. When those characters appear elsewhere (for example, in the value of a parameter), they should be encoded as defined in IETF RFC 2396. The server shall be prepared to decode any character escaped in this manner.

Table 1 — Reserved characters in SSA query string

| Character | Reserved usage |
|---|---|
| ? | Separator indicating start of query string. |
| & | Separator between parameters in query string. |
| = | Separator between name and value of parameter. |
| ,/; | Separator between individual values in list-oriented parameters (such as POS, BAND, TIME, etc.). |

In particular, if any parameter value contains the character "#" (for example in a dataset identifier) it must be URL encoded to be legally included in a URL.

### 8.3.3 HTTP GET

A SSA service shall support the "GET" method of the HTTP protocol (IETF RFC 2616).

An Online Resource URL intended for HTTP GET requests is in fact only a URL prefix to which additional parameters are appended in order to construct a valid Operation request. A URL prefix is defined in accordance with IETF RFC 2396 as a string including, in order, the scheme ("http" or "https"), Internet Protocol hostname or numeric address, optional port number, path, mandatory question mark "?", and optional string comprising one or more server-specific parameters ending in an ampersand "&". The prefix defines the network address to which request messages are to be sent for a particular operation on a particular server. Each operation may have a different prefix. Each prefix is entirely at the discretion of the service provider.

This document defines how to construct a query part that is appended to the URL prefix in order to form a complete request message. Every SSA operation has several mandatory or optional request parameters. Each parameter has a defined name. Each parameter may have one or more legal values, which are either defined by this document or are selected by the client based on service metadata. To formulate the query part of the URL, a client shall append the mandatory request parameters, and any desired optional parameters, as name/value pairs in the form "name=value&" (parameter name, equals sign, parameter value, ampersand). The "&" is a separator between name/value pairs, and is therefore optional after the last pair in the request string.

When the HTTP GET method is used, the client-constructed query part is appended to the URL prefix defined by the server, and the resulting complete URL is invoked as defined by HTTP (IETF RFC 2616).

Table 2 summarizes the components of an operation request URL when HTTP GET is used.

Table 2 — Structure of SSA request using HTTP GET

| URL component | Description |
|---|---|
| http://host:port]/path[?[name[=value]{&name=[value]}]] | Base-URL (prefix) of service operation. [] denotes 0 or 1 occurrence of an optional part; {} denotes 0 or more occurences. |
| name=value& | One or more standard request parameter name/value pairs as defined for each operation by this document. |

### 8.3.4 HTTP POST

SSA does not currently support the "POST" method of the HTTP protocol (IETF RFC 2616), but may do so in the future. POST could be used, for example, to permit large range-lists to be specified.

## 8.4 General HTTP response rules

Upon receiving a valid request, the server shall send a response corresponding exactly to the request as detailed in section 4.2 of this document, or send a service exception if unable to respond correctly. Only in the case of Version Negotiation (see 8.2.4) may the server offer a differing result. Upon receiving an invalid request, the server shall issue a service exception as described in 8.10.

A server may send an HTTP Redirect message (using HTTP response codes as defined in IETF RFC 2616) to an absolute URL that is different from the valid request URL that was sent by the client. HTTP Redirect causes the client to issue a new HTTP request for the new

URL. Several redirects could in theory occur. Practically speaking, the redirect sequence ends when the server responds with a SSA response. The final response shall be a SSA response that corresponds exactly to the original request (or a service exception).

Response objects shall be accompanied by the appropriate Multipurpose Internet Mail Extensions (MIME) type (IETF RFC 2045) for that object. A list of MIME types in common use on the internet is maintained by the Internet Assigned Numbers Authority (IANA). Allowable types for operation responses and service exceptions are discussed below. The basic structure of a MIME type is a string of the form "type/subtype". MIME allows additional parameters in a string of the form "type/subtype; param1=value1; param2=value2". A server may include parameterized MIME types in its list of supported output formats. In addition to any parameterized variants, the server should offer the basic unparameterized version of the format.

Response objects should be accompanied by other HTTP entity headers as appropriate and to the extent possible. In particular, the Expires and Last-Modified headers provide important information for caching; Content-Length may be used by clients to know when data transmission is complete and to efficiently allocate space for results, and Content-Encoding or Content-Transfer-Encoding may be necessary for proper interpretation of the results.

## 8.5 Numeric and boolean values

Integer numbers shall be represented in a manner consistent with the specification for integers in XML Schema Datatypes. This document shall explicitly indicate where an integer value is mandatory.  Real numbers shall be represented in a manner consistent with the specification for double-precision numbers in XML Schema Datatypes. This representation allows for integer, decimal and exponential notations. A real value is allowed in all numeric fields defined by this document unless the value is explicitly restricted to integer.

Sexagesimal formatting is not permitted other than in ISO 8601 formatted time strings unless otherwise specified in this document.  In particular, astronomical coordinates should be rendered as real numbers as specified above.

Positive, negative and zero values are allowed unless explicitly restricted.

Boolean values shall be represented in a manner consistent with the specification for Boolean in XML Schema Datatypes. The values "0" and "false" are equivalent. The values "1" and "true" are equivalent. Absence of an optional value is equivalent to logical false. This document shall explicitly indicate where a Boolean value is mandatory.

## 8.6 Output formats

The response to a SSA request is always a computer file that is transferred over the Internet from the server to the client. The file may contain text, or the file may be a graphics or FITS-formatted file. As stated in 7.2, the type of the returned file shall be indicated by a MIME type string.

Text output formats are usually formatted as Extensible Markup Language (XML; MIME type text/xml). Text formats are used to convey service metadata, descriptions of error conditions, or responses to data queries.  In particular, the response to a  data query is always returned as an XML file in VOTable format.

## *8.7 Request parameter rules*

### 8.7.1 Parameter ordering and case

Parameter names shall not be case sensitive, but parameter values shall be. In this document, parameter names are typically shown in uppercase for typographical clarity, not as a requirement.

Parameters in a request may be specified in any order.

When request parameters are duplicated with conflicting values, the response from the server may be undefined. This document does not mandate which of the duplicated values sent by the client are to be used by the server. It is recommended that neither the client nor the service should repeat parameter values in a query URL.

A SSA service shall be prepared to encounter additional request parameters that are not part of this document without reporting an error. In terms of producing results per this document, a SSA service shall not require such parameters, but may define additional service-defined parameters.

### 8.7.2 Range-list parameters

Parameters which are *list-valued* (for example, POS, BAND and TIME) use the comma (",") as the separator between successive items in the list. Embedded white space is not permitted. If a parameter value includes a space or comma, it must be escaped using the URL encoding rules (see 8.3.2 and IETF RFC 2396).

In some lists, individual entries may be empty, and should be represented by the empty string. Thus, two successive commas indicate an empty item, as does a leading comma or a trailing comma. An empty list should be interpreted either as a list containing no items, or as a list containing a single empty item, depending upon the context.

Some parameters (for example BAND and TIME) may allow a parameter value to be specified as a numeric range. Such *range-valued* parameters use the forward slash ("/") character as the separator between elements of the range specification (as in the ISO 8601 date specification after which this convention is patterned). For example, "5E-7/8E-7" would specify a range consisting of all values from 5E-7 to 8E-7, inclusive. If a third field is specified it is a step size for traversing the indicated range. If a parameter permits a step size the semantics of the step size are defined by the specific parameter.

An open range may be specified by omitting either range value. If the first value is omitted the range is open toward *lower* values. If the second value is omitted the range is open toward *higher* values. Omitting both values indicates an infinite range which accepts all values. For example, "/5" is an open range which accepts all values less than or equal to 5. To specify all values less than 5, "/4" would be used (for an integer valued range). Range values are limited to numeric values or ISO dates.

A list may be qualified by appending the character ";" (semicolon) followed by a qualifier string. For example "1E-7/3E-6;source" could specify a spectral bandpass in the rest frame of the source.

List and range syntax may be combined, e.g., to indicate a list of scalar or range-valued parameter values.  Such a *range list* may be ordered or unordered, and may contain either numeric or string data.  An *ordered* list is one which requires values to be processed in a specified order, and to ensure this the range list is sorted or ordered *by the service* as necessary before being used.  It is the responsibility of the service to sort an ordered range list, hence the client can input ranges or range values in any order for an ordered range list and the result will be the same.  The sequence in which items in an *unordered* list occur on the other hand is significant, as since there is no intrinsic ordering for the list which can be enforced by the service, items will be processed by the service in the order they are input by the client.

TIME and BAND are typical examples of ordered range lists.  Since a dataset matches the query if it contains data in any of the specified ranges, logically it does not matter in what order the ranges are given, or whether the first element of a range is less than the second, or whether ranges overlap; the result should be the same in all cases.  Hence the range list has an intrinsic ordering irrespective of how ranges are input.  Unless otherwise specified in the definition of a given parameter, range lists are assumed to be ordered.

### 8.7.3 Missing or null-valued parameters

If a parameter is not included in a query its value is *unset*; no value has been specified.  If a parameter is given a null value, e.g., "POS=", the parameter value has been set and the value is the null string.  The interpretation of such an input is defined separately for each parameter, and may or may not be an error condition.

## 8.8  Common request parameters

### 8.8.1 VERSION

The VERSION parameter specifies the protocol version number. The format of the version number, and version negotiation, are described in 8.2.

### 8.8.2 REQUEST

The REQUEST parameter indicates which service operation is being invoked. The value shall be the name of one of the operations offered by the server.  It is an error to reference an unknown service operation.

### 8.8.3 Extended capabilities and operations

The SSA service allows for optional extended capabilities and operations. Extensions may be defined within an information community when needed for additional functionality or specialization. A generic client shall not be required or expected to make use of such extensions. Extended capabilities or operations shall be defined by the service metadata. Extended capabilities provide additional metadata about the service, and may or may not enable optional new parameters to be included in operation requests. Extended operations allow additional operations to be defined.

A server shall produce a valid response to the operations defined in this document, even if parameters used by extended capabilities are missing or malformed (*i.e.* the server shall supply a default value for any extended capabilities it defines), or if parameters are supplied that are not known to the server.

Service providers shall choose extension names with care to avoid conflicting with standard metadata fields, parameters and operations.

## 8.9 Service result

The return value of a valid Service request shall correspond to the output type specified for the operation, or requested in the FORMAT parameter in the case of an operation which can return data in a choice of output formats. In an HTTP environment, the Content-type header of the response shall be exactly the MIME type associated with the valid request.

## 8.10 Error Response and Other Unsuccessful Results

Upon receiving a request that is invalid according to this document, the server shall issue a service exception report. The service exception report is meant to describe to the client application or its human user the reason(s) that the request is invalid. The allowed service exception formats are defined below.

If a service operation throws an error response and exits, the default action of the service should be to return a VOTable noting that an error has occurred, and describing the error. An INFO element within the "results" RESOURCE element of the VOTable is used to indicate success or failure of the operation. As described in the previous section, the INFO element must have `name="QUERY_STATUS"`; if the operation is successful (regardless of whether any data is returned) the value attribute is set to `"OK"`. The remainder of this section defines other possible values to indicate that the query was unsuccessful in some way. When the query is unsuccessful, the contents of INFO element (i.e. its PCDATA child node) **should** contain an error message suitable for display.

When the query is unsuccessful (in any of the senses described below), the resulting VOTable is not required to contain any other elements as specified for a successful operation; however, it is not an error to do so. For example, additional INFO elements may be returned to echo back the input parameters of the operation which failed, as in the following example.

**Example:**
```
<VOTABLE … version="1.1">
  <RESOURCE type="results">
    <INFO name="QUERY_STATUS" value="ERROR">unrecognized operation</INFO>
    <INFO name="SERVICE_PROTOCOL" value="1.0">SSAP</INFO>

    <INFO name="REQUEST" value="queryData"/>
    <INFO name="baseUrl" value="http://webtest.aoc.nrao.edu/ivoa-dal"/>
    <INFO name="serviceVersion" value="1.0"/>
    <INFO name="serviceName" value="ssap"/>
    <INFO name="ServiceEngine" value="ssap: SSAP 1.0 DALServer version 0.1"/>
  </RESOURCE>
</VOTABLE>
```

The other allowed values for the value attribute besides `OK` are as specified below.

### 8.10.1  Service Error

The server failed to process the operation.  Typical reasons include:

- The input query contained a syntax error.
- The way the query was posed was invalid for some reason, e.g., due to an invalid query region specification.
- A constraint parameter value was given an illegal value; e.g. `DEC=91`.
- The server trapped an internal error (e.g., failed to connect to its database) preventing further processing.

In this case, the inclusion of a descriptive error message **should** be returned.

### 8.10.2  Overflow

Overflow indicates that the operation produced results that exceeded the limits of the service in some way.  For instance, a data query matched too many candidate datasets, exceeding the current allowable maximum number of output records.  In this case, the service **should** include an error message indicating the nature of the overflow condition.

```
Example:
    <INFO name="QUERY_STATUS" value="ERROR">DEC out of range: DEC=91</INFO>
    <INFO name="QUERY_STATUS" value="OVERFLOW">Number of matching spectra
    exceeds default limit of 500</INFO>
```

If overflow occurs a query status of OVERFLOW **must** be indicated, returning an otherwise valid query response containing the maximum number of records.  The query may be repeated, requesting a higher MAXREC value than the default for the maximum number of output records, up to the hard limit defined by the service capabilities.  Alternatively the query parameters may be adjusted to more carefully constrain the query.  Currently these are the only ways to avoid overflow when performing a query.

### 8.10.3 Other Errors

Although the intention is that service should catch all errors and return a uniform error response in the prescribed VOTable format, informing the client of the nature of the error which occurred in service-specific terms, this is not always possible.  More fundamental errors may result in a HTTP level error.  The client should be prepared to handle either form of error.  Which is returned in a given case, may depend upon the operation performed, the nature of the error, and the details of how a given service is implemented.

## 9   Changes from Previous Versions

### 9.1 Version 1.0 to 1.1

- Clarified relationship of SSA and SDM data models.
- Eliminated use of wildcards to compress long Utypes.
- Fixed UCD for spectral axis coverage fields.
- Deleted "Data" fields from data model summary since these do not apply to the SSA query response, only to a spectral dataset.

- In the discussion of overflow of the query response added a clarification stating that if overflow occurs an otherwise valid query response containing the maximum number of records should be returned, with query status set to OVERFLOW.
- Modified the description of ISO8601 (time format) usage to restrict the allowable formats to a reasonable subset, consistent with other IVOA standards.
- Added the appropriate IVOA Architecture figure for SSA.
- Added a "changes from previous versions" section.
- Updated the References section.
- Changes from RFC including various UCD updates re S.Derriere.

# Appendix A: Theoretical Spectral Access Use Case

Theoretical models are widely used in Astronomy. The physical properties of an object can, for instance, be inferred by comparing its observed spectrum to a grid of theoretical spectra. Theoretical models are usually available in the Internet as collection of data files that can be downloaded (in some case with the help of a Web form allowing a previous selection of the files of interest) in different formats like ASCII or FITS. Building an extra layer on top of these services implementing the SSAP protocol allows for the seamless sharing of theoretical models within the VO community.

This paragraph describes the usage of the FORMAT=METADATA mechanism to access Theoretical Spectra in the context of the Simple Spectral Access Protocol.

### The "Client-Server Parameter" dialogue

The usage of the SSAP to access gridded models of theoretical spectra can be described as a dialogue between the client application and the model server based in three main steps:

1. The client makes a request using the FORMAT=METADATA operation to get the list of available parameters for the model set
2. The client (whether automatically or by human intervention) makes a selection of the desired parameters/ranges for the required models and sends a request for data to the server
3. The server responds with a VOTable containing metadata pertaining to the specified parameter models plus the corresponding "access reference" to the model data.

Note that the request of POS-SIZE, BAND or TIME are not required for this dialogue.

The following example uses a general SSAP service mounted on a generic URL like the following http://modelserver.com/ssap.php.

### Step by step process

The client (VO application, Web interface, etc) sends an HTTP query to the server asking for metadata describing what the server offers (in the future this will use *getCapabilities*):

[http://modelserver.com/ssap.php?REQUEST=queryData&FORMAT=metadata](http://modelserver.com/ssap.php?REQUEST=queryData&FORMAT=metadata)

The server answers the query by sending a VOTable containing information about the server itself, the list of parameters allowed in the query, their description and, optionally, the accepted values or ranges of values:

```
<description> a general text description of the theoretical model
     offered by the server </description>
  <param name="param1" ucd="..." datatype="float"...>
    <description> a short human-readable text description of the
        meaning of this parameter</description>
  </param>
     ...
```

For instance, two parameters like Effective Temperature and Surface Gravity could be described as follows:

```
<description>Stellar atmosphere model by..., version 1 </description>
<param name="teff" ucd="phys.temperature.effective" units="K"
   datatype="float">
   <description>Effective temperature for the model in K</description>
</param>
<param name="logg" ucd="phys.gravity" datatype="float">
   <description>Logarithm of the surface gravity</description>
</param>
```

The client, reading this VOTable response, knows the parameters –and their ranges- available for the search as well as their names and descriptions and can build a small user interface on-the-fly, like the following one (in this case, for the following parameter ranges: Effective Temperature, Surface Gravity and Metallicity):

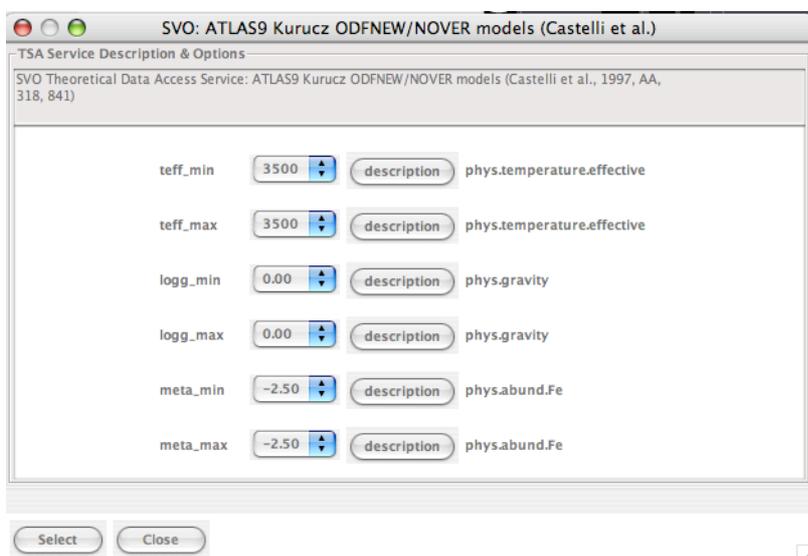

When the user makes the selection, the client sends a search query to the server with the constraints on the parameters:

http://modelserver.com/ssap.php?param1=value1&...¶mN=valueN

The server answers the query with a VOTable containing a list with all the particular instances of the model that are available within the specified search criteria. One of the expected fields is the Access.Reference that allows the model spectrum to be retrieved.

Examples of valid VOTable responses for the FORMAT=METADATA operation and for the queryData operation are given here for reference:

### *Sample VOTable response for a Theory-SSAP FORMAT=METADATA request*

Note that ellipsis symbol between brackets "[…]" substitutes output of values for brevity.

```
<VOTABLE version="1.1" xsi:schemaLocation="http://www.ivoa.net/xml/VOTable/v1.1">

<RESOURCE type="meta">
       <DESCRIPTION>
              SVO Theoretical Data Access Service: ATLAS9 Kurucz ODFNEW/NOVER models
              (Castelli et al., 1997, AA, 318, 841)
       </DESCRIPTION>
<INFO name="QUERY_STATUS" value="OK">
<PARAM name="INPUT:teff_min" ucd="phys.temperature.effective">
       <DESCRIPTION>
              min value for the effective temperature for the model. Temperatures are given
       in K
       </DESCRIPTION>
       <VALUES type="actual">
              <OPTION value="3500"/>
              <OPTION value="3750"/>
              <OPTION value="4000"/>
              […]
              <OPTION value="49000"/>
              <OPTION value="50000"/>
       </VALUES>
</PARAM>
<PARAM name="INPUT:teff_max" ucd="phys.temperature.effective">
       <DESCRIPTION>
              max value for the effective temperature for the model. Temperatures are given
       in K
       </DESCRIPTION>
       <VALUES type="actual">
              <OPTION value="3500"/>
              <OPTION value="3750"/>
              […]
              <OPTION value="48000"/>
              <OPTION value="49000"/>
              <OPTION value="50000"/>
       </VALUES>
</PARAM>
<PARAM name="INPUT:logg_min" ucd="phys.gravity">
       <DESCRIPTION>min value for Log(G) for the model.</DESCRIPTION>
       <VALUES type="actual">
              <OPTION value="0.00"/>
              […]
              <OPTION value="4.50"/>
              <OPTION value="5.00"/>
       </VALUES>
</PARAM>
<PARAM name="INPUT:logg_max" ucd="phys.gravity">
       <DESCRIPTION>max value for Log(G) for the model.</DESCRIPTION>
```

```
        <VALUES type="actual">
                <OPTION value="0.00"/>
                […]
                <OPTION value="4.50"/>
                <OPTION value="5.00"/>
        </VALUES>
</PARAM>
<PARAM name="INPUT:meta_min" ucd="phys.abund.Fe">
        <DESCRIPTION>min value for the Metallicity for the model.</DESCRIPTION>
        <VALUES type="actual">
                <OPTION value="-2.50"/>
                […]
                <OPTION value="0.20"/>
                <OPTION value="0.50"/>
        </VALUES>
</PARAM>
<PARAM name="INPUT:meta_max" ucd="phys.abund.Fe">
<DESCRIPTION>max value for the Metallicity for the model.</DESCRIPTION>
        <VALUES type="actual">
                <OPTION value="-2.50"/>
                <OPTION value="-2.00"/>
                […]
                <OPTION value="0.20"/>
                <OPTION value="0.50"/>
        </VALUES>
</PARAM>

<TABLE>
        <DESCRIPTION>
                ODFNEW /NOVER models. Newly computed ODFs with better opacities and better
                abundances have been used.
        </DESCRIPTION>
<PARAM name="DataModel" utype="ssa:Dataset.DataModel" datatype="char" arraysize="*"
value="Spectrum 1.0">
        <DESCRIPTION>Data Model name and version</DESCRIPTION>
</PARAM>
<PARAM name="Publisher" utype="ssa:Curation.Publisher" ucd="meta.curation" datatype="char"
arraysize="*" value="LAEFF/Spanish Virtual Observatory"/>
<PARAM name="PubID" utype="ssa:Curation.PublisherID" ucd="meta.ref.url;meta.curation"
datatype="char" arraysize="*" value="TBD"/>

<PARAM name="Creator" utype="ssa:DataID.Creator" ucd="" datatype="char" arraysize="*"
value="LAEFF/Spanish Virtual Observatory"/>
<PARAM name="Collection" utype="ssa:DataID.Collection" ucd="" datatype="char"
arraysize="*" value="Kurucz models of stellar atmposheres"/>
<PARAM name="CreationType" utype="ssa:DataID.CreationType" ucd="" datatype="char"
arraysize="*" value="Simulation"/>

<FIELD name="teff" ucd="phys.temperature.effective" unit="K" datatype="int">
        <DESCRIPTION>
        value for the effective temperature for the model. Temperatures are given in K
        </DESCRIPTION>
</FIELD>
<FIELD name="logg" ucd="phys.gravity" unit="log(cm/s**2)" datatype="float">
        <DESCRIPTION>value for Log(G) for the model.</DESCRIPTION>
</FIELD>
<FIELD name="meta" ucd="phys.abund.Fe" unit="" datatype="float">
        <DESCRIPTION>value for the Metallicity for the model.</DESCRIPTION>
</FIELD>
<FIELD name="vtur" ucd="phys.veloc.microTurb" unit="km/s" datatype="float">
        <DESCRIPTION>Microturbulence velocity</DESCRIPTION>
</FIELD>
<FIELD name="lh" ucd="VOX:lh" unit="" datatype="float">
        <DESCRIPTION>
                l/Hp where l is the  mixing length of the convective element and Hp is the
                pressure scale height
        </DESCRIPTION>
```

```
</FIELD>

<FIELD name="title" ucd="meta.title;meta.dataset" utype="ssa:DataId.Title" datatype="char"
arraysize="*">
        <DESCRIPTION>Title.</DESCRIPTION>
</FIELD>
<FIELD name="SpectralAxis" utype="ssa:Dataset.SpectralAxis" datatype="char" arraysize="*">
        <DESCRIPTION>
                Flux Axis name.
        </DESCRIPTION>
</FIELD>
<FIELD name="FluxAxis" utype="ssa:Dataset.FluxAsis" datatype="char" arraysize="*">
        <DESCRIPTION>
                Spectral Axis name.
        </DESCRIPTION>
</FIELD>
<FIELD name="SpectralSI" utype="ssa:Dataset.SpectralSI" datatype="char" arraysize="*">
        <DESCRIPTION>
                SpectralAxis SI conversion factor and dimentions (blank separated).
                E.g.,
                10-10 L
                to imply Angstrom, in the International System of Units.
        </DESCRIPTION>
</FIELD>
<FIELD name="FluxSI" utype="ssa:Dataset.FluxSI" datatype="char" arraysize="*">
        <DESCRIPTION>
                Flux Axis SI conversion factor and dimensions (blank separated).
                E.g.
                10+7 ML-1T-3
                to imply erg/cm2/sec/A in the International System of Units.
        </DESCRIPTION>
</FIELD>
<FIELD name="UNITS" ucd="meta.unit" datatype="char" arraysize="*">
        <DESCRIPTION>
                Units in each of the axes.
        </DESCRIPTION>
</FIELD>
<FIELD name="DataLength" utype="ssa:Dataset.Length" datatype="char" arraysize="*">
        <DESCRIPTION>Number of points</DESCRIPTION>
</FIELD>
<FIELD name="format" utype="ssa:Access.Format" datatype="char" arraysize="*">
        <DESCRIPTION>Spectrum format</DESCRIPTION>
</FIELD>
<FIELD name="Spectrum" utype="ssa:Access.Reference"  datatype="char" arraysize="*">
        <DESCRIPTION>Link to the spectrum file</DESCRIPTION>
</FIELD>

</TABLE>
</RESOURCE>
</VOTABLE>
```

### Sample VOTable response for a Theory-SSAP queryData request

Sample query:

http://svo.laeff.inta.es/projects/svo/theory/db2vo/html/tsap.p
hp?REQUEST=queryData&model=kurucz&teff_min=3500&teff_max=3500&
logg_min=0.00&logg_max=0.5&meta_min=-2.5&meta_max=-2.0

The following output is returned:

```
<VOTABLE version="1.1" xsi:schemaLocation="http://www.ivoa.net/xml/VOTable/v1.1"
```

```
        xmlns:ssa="http://www.ivoa.net/xml/DalSsap/v1.0">
<RESOURCE type="results">
<DESCRIPTION>
        SVO Theoretical Data Access Service: ATLAS9 Kurucz ODFNEW/NOVER models (Castelli et
        al., 1997, AA, 318, 841)
</DESCRIPTION>
<INFO name="QUERY_STATUS" value="OK"/>
<TABLE>
<DESCRIPTION>dalessio models. Search results.</DESCRIPTION>
<PARAM name="DataModel" utype="ssa:Dataset.DataModel" datatype="char" arraysize="*"
value="Spectrum 1.0">
        <DESCRIPTION>Data Model name and version</DESCRIPTION>
</PARAM>

<PARAM name="Publisher" utype="ssa:Curation.Publisher" ucd=" meta.curation"
datatype="char" arraysize="*" value="LAEFF/Spanish Virtual Observatory"/>
<PARAM name="PubID" utype="ssa:Curation.PublisherID" ucd="meta.ref.url;meta.curation"
datatype="char" arraysize="*" value="TBD"/>

<PARAM name="Creator" utype="ssa:DataID.Creator" ucd="" datatype="char" arraysize="*"
value="LAEFF/Spanish Virtual Observatory"/>
<PARAM name="Collection" utype="ssa:DataID.Collection" ucd="" datatype="char"
arraysize="*" value="TBD">
        <DESCRIPTION>Kurucz models of stellar atmopsheres</DESCRIPTION>
</PARAM>
<PARAM name="DataSource" utype="ssa:DataID.DataSource" ucd="" datatype="char"
arraysize="*" value="Theory"/>
<PARAM name="CreationType" utype="ssa:DataID.CreationType" ucd="" datatype="char"
arraysize="*" value="Archival"/>

<FIELD name="teff" ucd="phys.temperature.effective" unit="K" datatype="int">
        <DESCRIPTION>
        value for the effective temperature for the model. Temperatures are given in K
        </DESCRIPTION>
</FIELD>
<FIELD name="logg" ucd="phys.gravity" unit="log(cm/s**2)" datatype="float">
        <DESCRIPTION>value for Log(G) for the model.</DESCRIPTION>
</FIELD>
<FIELD name="meta" ucd="phys.abund.Fe" unit="" datatype="float">
        <DESCRIPTION>value for the Metallicity for the model.</DESCRIPTION>
</FIELD>
<FIELD name="vtur" ucd="phys.veloc.microTurb" unit="km/s" datatype="float">
        <DESCRIPTION>Microturbulence velocity</DESCRIPTION>
</FIELD>
<FIELD name="lh" ucd="VOX:lh" unit="" datatype="float">
        <DESCRIPTION>
                l/Hp where l is the  mixing length of the convective element and Hp is the
                pressure scale height
        </DESCRIPTION>
</FIELD>

<FIELD name="title" ucd="meta.title;meta.dataset" utype="ssa:DataId.Title" datatype="char"
arraysize="*">
        <DESCRIPTION>Title.</DESCRIPTION>
</FIELD>
<FIELD name="SpectralAxis" utype="Dataset.SpectralAxis" datatype="char" arraysize="*">
        <DESCRIPTION>
                Spectral Axis name.
        </DESCRIPTION>
</FIELD>
<FIELD name="FluxAxis" utype="Dataset.FluxAxis" datatype="char" arraysize="*">
        <DESCRIPTION>
                Flux Axis name.
        </DESCRIPTION>
</FIELD>
<FIELD name="SpectralSI" utype="Dataset.SpectralSI" datatype="char" arraysize="*">
        <DESCRIPTION>
```

```
            SpectralAxis SI conversion factor and dimentions (blank separated).
            E.g.,
            10-10 L
            to imply Angstrom, in the International System of Units.
        </DESCRIPTION>
</FIELD>
<FIELD name="FluxSI" utype="Dataset.FluxSI" datatype="char" arraysize="*">
        <DESCRIPTION> Flux Axis SI conversion factor and dimensions (blank separated).
            E.g.
            10+7 ML-1T-3
            to imply erg/cm2/sec/A in the International System of Units.
        </DESCRIPTION>
</FIELD>
<FIELD name="UNITS" ucd="meta.unit" datatype="char" arraysize="*">
        <DESCRIPTION>Units in each of the axes.</DESCRIPTION>
</FIELD>
<FIELD name="DataLength" utype="ssa:Dataset.Length" datatype="char" arraysize="*">
        <DESCRIPTION>Number of points</DESCRIPTION>
</FIELD>
<FIELD name="format" utype="ssa:Access.Format" datatype="char" arraysize="*">
        <DESCRIPTION>Spectrum format</DESCRIPTION>
</FIELD>
<FIELD name="Spectrum" utype="ssa:Access.Reference"  datatype="char" arraysize="*">
        <DESCRIPTION>Link to the spectrum file</DESCRIPTION>
</FIELD>
<DATA>
<TABLEDATA>
<TR>
        <TD>3500</TD>
        <TD>0.00</TD>
        <TD>-2.00</TD>
        <TD>2.0</TD>
        <TD>1.25</TD>
        <TD>Kurucz ODFNEW /NOVER, teff:3500, logg:0.00, meta:-2.00</TD>
        <TD>WAVELENGTH</TD>
        <TD>FLUX</TD>
        <TD>10-10 L</TD>
        <TD>10+7 MT-1L-2</TD>
        <TD>ANGSTROM ERG/CM2/S/A</TD>
        <TD>1221</TD>
        <TD>application/x-votable+xml</TD>
        <TD>
http://svo.laeff.inta.es/projects/svo/theory/db2vo/html/tsap.php?model=Kurucz&id=963</
TD>
</TR>
<TR>
        <TD>3500</TD>
        <TD>0.50</TD>
        <TD>-2.00</TD>
        <TD>2.0</TD>
        <TD>1.25</TD>
        <TD>Kurucz ODFNEW /NOVER, teff:3500, logg:0.50, meta:-2.00</TD>
        <TD>WAVELENGTH</TD>
        <TD>FLUX</TD>
        <TD>10-10 L</TD>
        <TD>10+7 ML-1T-3</TD>
        <TD>ANGSTROM ERG/CM2/S/A</TD>
        <TD>1221</TD>
        <TD>application/x-votable+xml</TD>
        <TD>
http://svo.laeff.inta.es/projects/svo/theory/db2vo/html/tsap.php?model=Kurucz&id=964
</TD>
</TR>
        […]
</TABLEDATA>
</DATA>
</TABLE>
</RESOURCE>
```

```
</VOTABLE>
```

# Appendix B: Standard QueryData Query Response

The output below illustrates the query response from a working SSA service.  In the interests of brevity some of the FIELD and GROUP definitions have been omitted, and only data for a single table row is included (this output was computer generated and minimal effort has been made to pretty-print it):

```
<VOTABLE xmlns:xsi="http://www.w3.org/2001/XMLSchema-instance"
 xsi:noNamespaceSchemaLocation="xmlns:http://www.ivoa.net/xml/VOTable/VOTable-1.1.xsd"
 xmlns:ssa="http://www.ivoa.net/xml/DalSsap/v1.0" version="1.1">

<RESOURCE type="Results">
<DESCRIPTION>DALServer proxy service for JHU spectrum services</DESCRIPTION>
<INFO name="QUERY_STATUS" value="OK"/>
<INFO name="SERVICE_PROTOCOL" value="1.0">SSAP</INFO>

<INFO name="REQUEST" value="queryData"/>
<INFO name="POS" value="180.0,1.0"/>
<INFO name="SIZE" value="0.2"/>
<INFO name="FORMAT" value="all"/>
<INFO name="Collection" value="ivo://jhu/sdss/dr5"/>
<INFO name="ServiceEngine" value="JhuProxySsap: SSAP 1.0 DALServer version 0.1"/>
<INFO name="TableRows" value="42"/>
<TABLE>
<FIELD ID="Score" name="Score" datatype="float" utype="ssa:Query.Score">
    <DESCRIPTION>Degree of match to query parameters</DESCRIPTION>
</FIELD>
<FIELD ID="AssocID" name="AssocID" datatype="char" utype="ssa:Association.ID"
    arraysize="*">
    <DESCRIPTION>Association identifier</DESCRIPTION>
</FIELD>
<FIELD ID="AcRef" name="AcRef" datatype="char" ucd="meta.ref.url"
     utype="ssa:Access.Reference" arraysize="*">
    <DESCRIPTION>URL used to access dataset</DESCRIPTION>
</FIELD>
<FIELD ID="Format" name="Format" datatype="char" utype="ssa:Access.Format"
    arraysize="*">
    <DESCRIPTION>Content or MIME type of dataset</DESCRIPTION>
</FIELD>
<FIELD ID="DataModel" name="DataModel" datatype="char" utype="ssa:Dataset.DataModel"
     arraysize="*">
    <DESCRIPTION>Datamodel name and version</DESCRIPTION>
</FIELD>
<FIELD ID="DataLength" name="DataLength" datatype="long" utype="ssa:Dataset.Length">
    <DESCRIPTION>Number of points</DESCRIPTION>
</FIELD>
<FIELD ID="Title" name="Title" datatype="char" ucd="meta.title;meta.dataset"
     utype="ssa:DataID.Title" arraysize="*">
<DESCRIPTION>Dataset Title</DESCRIPTION>
</FIELD>
<FIELD ID="Creator" name="Creator" datatype="char" utype="ssa:DataID.Creator"
     arraysize="*">
<DESCRIPTION>Dataset creator</DESCRIPTION>
</FIELD>
<FIELD ID="Collection" name="Collection" datatype="char" utype="ssa:DataID.Collection"
     arraysize="*">
    <DESCRIPTION>Data collection to which dataset belongs</DESCRIPTION>
</FIELD>
<FIELD ID="CreatorDID" name="CreatorDID" datatype="char" ucd="meta.id"
     utype="ssa:DataID.CreatorDID" arraysize="*">
    <DESCRIPTION>Creator's ID for the dataset</DESCRIPTION>
</FIELD>
<FIELD ID="CreatorDate" name="CreatorDate" datatype="char" ucd="time;meta.dataset"
```

```
      utype="ssa:DataID.Date" arraysize="*">
     <DESCRIPTION>Data processing/creation date</DESCRIPTION>
</FIELD>
     [more FIELDs ommitted]

<GROUP ID="Query" name="Query" utype="ssa:Query">
     <DESCRIPTION>Query Metadata</DESCRIPTION>
     <FIELDref ref="Score"/>
</GROUP>
<GROUP ID="Association" name="Association" utype="ssa:Association">
     <DESCRIPTION>Association Metadata</DESCRIPTION>
     <FIELDref ref="AssocID"/>
     <PARAM ID="AssocType" datatype="char" name="AssocType"
          utype="ssa:Association.Type" value="MultiFormat" arraysize="*">
          <DESCRIPTION>Type of association</DESCRIPTION>
     </PARAM>
     <PARAM ID="AssocKey" datatype="char" name="AssocKey" utype="ssa:Association.Key"
          value="@Format" arraysize="*">
          <DESCRIPTION>Key used to distinguish association elements</DESCRIPTION>
     </PARAM>
</GROUP>
<GROUP ID="Access" name="Access" utype="ssa:Access">
     <DESCRIPTION>Access Metadata</DESCRIPTION>
     <FIELDref ref="AcRef"/>
     <FIELDref ref="Format"/>
     <PARAM ID="DatasetSize" unit="byte" datatype="long" name="DatasetSize"
          utype="ssa:Access.Size" value="800000">
          <DESCRIPTION>Estimated dataset size</DESCRIPTION>
     </PARAM>
</GROUP>
<GROUP ID="Dataset" name="Dataset" utype="ssa:Dataset">
     <DESCRIPTION>General Dataset Metadata</DESCRIPTION>
     <FIELDref ref="DataModel"/>
     <FIELDref ref="DataLength"/>
     <PARAM ID="DatasetType" datatype="char" name="DatasetType"
          utype="ssa:Dataset.Type" value="Spectrum" arraysize="*">
          <DESCRIPTION>Dataset or segment type</DESCRIPTION>
     </PARAM>
</GROUP>
<GROUP ID="DataID" name="DataID" utype="ssa:DataID">
     <DESCRIPTION>Dataset Identification Metadata</DESCRIPTION>
     <FIELDref ref="Title"/>
     <FIELDref ref="Creator"/>
     <FIELDref ref="Collection"/>
     <FIELDref ref="CreatorDID"/>
     <FIELDref ref="CreatorDate"/>
     <FIELDref ref="CreatorVersion"/>
     <FIELDref ref="Instrument"/>
     PARAM ID="DataSource" datatype="char" name="DataSource"
          utype="ssa:DataID.DataSource" value="survey" arraysize="*">
          <DESCRIPTION>Original source of the data</DESCRIPTION>
     </PARAM>
     <PARAM ID="CreationType" datatype="char" name="CreationType"
          utype="ssa:DataID.CreationType" value="Archival" arraysize="*">
          <DESCRIPTION>Dataset creation type</DESCRIPTION>
     </PARAM>
</GROUP>
     [More GROUPs omitted]

<DATA>
<TABLEDATA>
     <TR>
          <TD>1.0</TD>
          <TD>MultiFormat.12</TD>
          <TD>http://webtest.aoc.nrao.edu/ivoa-dal/JhuProxySsap?
           REQUEST=getData&FORMAT=csv&
           PubDID=ivo%3A%2F%2Fjhu%2Fsdss%2Fdr5%2380442261170552832
          </TD>
```

```
        <TD>text/csv</TD>
        <TD>Spectrum 1.0</TD>
        <TD>4000</TD>
        <TD>SDSS J115923.80+000000.00 Galaxy 0285-51663-01</TD>
        <TD>sdss</TD>
        <TD>ivo://sdss/dr5/spec</TD>
        <TD>ivo://sdss/dr5/spec#80442261170552832</TD>
        <TD>2000-04-29T03:22:00.7900000-04:00</TD>
        <TD>3.13.1branch.1</TD>
        <TD>SDSS 2.5-M SPEC2 v4_5</TD>
        <TD>ivo://jhu/sdss/dr5#80442261170552832</TD>
        <TD>SDSS J115923.80+000000.00</TD>
        <TD>Galaxy</TD>
        <TD>0.451652</TD>
        <TD>0</TD>
        <TD>FK5</TD>
        <TD>2000</TD>
        <TD>TAI</TD>
        <TD>179.849160 .984768</TD>
        <TD>0.00083333333333333339</TD>
        <TD>em.wl</TD>
        <TD>6518.40990663334</TD>
        <TD>5389.1347401044286</TD>
        <TD>3823.8425365811263</TD>
        <TD>9212.9772766855549</TD>
        <TD>0</TD>
        <TD>0</TD>
        <TD>Absolute</TD>
        <TD>0</TD>
        <TD>51663.30695358796</TD>
        <TD>3600</TD>
        <TD>51663.2713056713</TD>
        <TD>51663.319500115744</TD>
        <TD>phot.fluDens;em.wl</TD>
        <TD>0</TD>
        <TD>Absolute</TD>
    </TR>
        [More table rows omitted]
</TABLEDATA>
</DATA>
</TABLE>
</RESOURCE>
</VOTABLE>
```

# Appendix C: Standard Metadata Query Response

The following example illustrates the response from a FORMAT=METADATA query, used to describe the parameters returned by the service instance.

```
<VOTABLE xmlns:xsi="http://www.w3.org/2001/XMLSchema-instance"
 xsi:noNamespaceSchemaLocation="xmlns:http://www.ivoa.net/xml/VOTable/VOTable-1.1.xsd"
 xmlns:ssa="http://www.ivoa.net/xml/DalSsap/v1.0" version="1.1">

<RESOURCE type="Results">
<DESCRIPTION>
        Sample of a getMetadata query response on a Simple Spectrum Access (SSA) service
</DESCRIPTION>

<INFO name="QUERY_STATUS" value="OK">Successful metadata query</INFO>
<INFO name="SERVICE_PROTOCOL" value="1.02">SSAP</INFO>

<!-- mandatory input parameters -->

    <PARAM name="INPUT:POS" value="" datatype="char" arraysize="*">
```

```
    <DESCRIPTION>
        The center of the region of interest.
        The coordinate values are specified in list format (comma separated) in
        decimal degrees with no embedded white space followed by an optional
        coord. systems such as GALACTIC_CENTER, TOPOCENTER, MARS; default is ICRS.
    </DESCRIPTION>
</PARAM>

<PARAM name="INPUT:SIZE" value="0.1" datatype="double" unit="deg">
    <DESCRIPTION>
        The radius of the circular region of interest in decimal degrees.
        A special case is SIZE=0. It will cause a search in the service defined
        default sized region of  0.1 degrees resulting in a patch of 0.01*pi sq.deg.
    </DESCRIPTION>
    <VALUES>
        <MIN value="0"/>
        <MAX value="5.0"/>
    </VALUES>
</PARAM>

<PARAM name="INPUT:BAND" value="ALL" datatype="char" arraysize="*">
    <DESCRIPTION>
        Spectral coverage: Several values can be combined in a
        comma separated list. Below values are treated case insensitive.
        All spectra returned by this service belong mainly to the optical
        reaching to the infrared regime. Therefore, the other values
        won't yield any matching records in the query response.
        Alternatively the wavelength can be given in meters or as a range thereof.
    </DESCRIPTION>
    <VALUES>
        <OPTION value="ALL"/>
        <OPTION value="radio"/>
        <OPTION value="millimeter"/>
        <OPTION value="infrared"/>
        <OPTION value="optical"/>
        <OPTION value="ultraviolet"/>
        <OPTION value="x-ray"/>
        <OPTION value="gamma-ray"/>
    </VALUES>
</PARAM>

<PARAM name="INPUT:TIME" value="" datatype="char" arraysize="*">
    <DESCRIPTION>
        If a single value is specified it matches any spectrum for which the time
        coverage includes the specified value.  If a range is specified it matches
        any spectrum which contains any data in the specified range.  Dates are
        expected in ISO 8601 UTC format.  E.g. 1998-05-21/1999-01-01 will search
        for all spectra taken in the given time period, that is starting 21st May,
        1998 to Jan 1st, 1999 inclusive.
    </DESCRIPTION>
</PARAM>

<PARAM name="INPUT:FORMAT" value="ALL" datatype="char" arraysize="*">
    <DESCRIPTION>
        Desired format of retrieved data.
        Note: The exact description of the output format
                (binary table or 1d image, definition of axes)
                is outside the scope of the access protocol.
                Below format values are treated case insensitive.</DESCRIPTION>
    <VALUES>
        <OPTION>ALL</OPTION>          <!-- search any format -->
        <OPTION>COMPLIANT</OPTION>  <!-- short for searching xml, votable, fits -->
        <OPTION>NATIVE</OPTION>      <!-- short for searching jpeg,png,legacy fits-->
        <OPTION>votable</OPTION>     <!-- short for application/x-votable+xml -->
        <OPTION>application/x-votable+xml</OPTION>
        <OPTION>fits</OPTION>          <!-- short for application/fits -->
        <OPTION>application/fits</OPTION>
        <OPTION>xml</OPTION>           <!-- short for application/xml -->
```

```
            <OPTION>application/xml</OPTION>
            <OPTION>GRAPHIC</OPTION>      <!-- short for searching jpeg and gif -->
            <OPTION>image/jpeg</OPTION>
            <OPTION>image/png</OPTION>
            <OPTION>METADATA</OPTION>
        </VALUES>
    </PARAM>

    <PARAM name="INPUT:REQUEST" datatype="char" arraysize="*">
        <DESCRIPTION>
            SSA protocol versions supported by this service.
            Reserved words for future extensions are:
            getData, stageData, getCapabilities, getAvailability
            Values are treated case-insensitive.
        </DESCRIPTION>
        <VALUES>
            <OPTION value="queryData" />
        </VALUES>
    </PARAM>

<!-- optional/recommended parameters and service defined input parameters -->

    <PARAM name="INPUT:VERSION" value="1.00" datatype="double">
        <DESCRIPTION>SSA protocol versions supported by this service.</DESCRIPTION>
        <VALUES>
            <OPTION value="1.0" />
            <OPTION value="1.02"/>
        </VALUES>
    </PARAM>

<!-- query response parameters (name="OUTPUT:param-name") -->

  <!-- service metadata -->

    <!-- service metadata: Query.* -->
    <PARAM ID="Score" name="OUTPUT:Score" datatype="float"
        utype="ssa:Query.Score" value="">
        <DESCRIPTION>Degree of match to query parameters</DESCRIPTION>
    </PARAM>
    <!-- service metadata: Access.* -->
    <PARAM ID="AcRef" name="OUTPUT:AcRef" datatype="char" ucd="meta.ref.url"
        utype="ssa:Access.Reference" arraysize="*" value="" >
        <DESCRIPTION>URL used to access dataset</DESCRIPTION>
    </PARAM>
    <PARAM ID="DisplayRef" name="OUTPUT:DisplayRef" datatype="char"
        ucd="meta.ref.url" utype="ssa:Access.Display" arraysize="*" value="" >
        <DESCRIPTION>URL used to display dataset</DESCRIPTION>
    </PARAM>

  <!-- data model metadata -->

    <!-- data model metadata: Dataset.* -->
    <PARAM ID="DataModel" name="OUTPUT:DataModel" datatype="char"
        utype="ssa:Dataset.DataModel" arraysize="*" value="" >
        <DESCRIPTION>Datamodel name and version</DESCRIPTION>
    </PARAM>
    <PARAM ID="DatasetType" datatype="char" name="OUTPUT:DatasetType"
        utype="ssa:Dataset.Type" value="Spectrum" arraysize="*" >
        <DESCRIPTION>Dataset or segment type</DESCRIPTION>
    </PARAM>

    <!-- data model metadata: DataID.* -->
    <PARAM ID="Title" name="OUTPUT:Title" datatype="char" ucd="meta.title;meta.dataset"
        utype="ssa:DataID.Title" arraysize="*" value="" >
        <DESCRIPTION>Dataset Title</DESCRIPTION>
    </PARAM>
    <PARAM ID="Creator" name="OUTPUT:Creator" datatype="char" utype="ssa:DataID.Creator"
        arraysize="*" value="" >
```

```
        <DESCRIPTION>Dataset creator</DESCRIPTION>
  </PARAM>
  <PARAM ID="Collection" name="OUTPUT:Collection" datatype="char"
      utype="ssa:DataID.Collection" arraysize="*" value="" >
      <DESCRIPTION>Data collection to which dataset belongs</DESCRIPTION>
  </PARAM>
  <PARAM ID="Instrument" name="OUTPUT:Instrument" datatype="char" ucd="meta.id;instr"
      utype="ssa:DataID.Instrument" arraysize="*" value="" >
      <DESCRIPTION>Instrument name</DESCRIPTION>
  </PARAM>
  <PARAM ID="CreatorDate" datatype="char" name="OUTPUT:CreatorDate"
      ucd="time;meta.dataset" utype="ssa:DataID.Date" arraysize="*" value="" >
      <DESCRIPTION>Data processing/creation date</DESCRIPTION>
  </PARAM>
  <PARAM ID="CreatorVersion" datatype="char" name="OUTPUT:CreatorVersion"
      ucd="meta.version;meta.dataset" utype="ssa:DataID.Version" arraysize="*"
      value="" >
      <DESCRIPTION>Version of dataset</DESCRIPTION>
  </PARAM>
  <PARAM ID="DataSource" datatype="char" name="OUTPUT:DataSource"
      utype="ssa:DataID.DataSource" value="Survey" arraysize="*" >
      <DESCRIPTION>Original source of the data</DESCRIPTION>
  </PARAM>
  <PARAM ID="CreationType" datatype="char" name="OUTPUT:CreationType"
      utype="ssa:DataID.CreationType" value="Archival" arraysize="*" >
      <DESCRIPTION>Dataset creation type</DESCRIPTION>
  </PARAM>

  <!-- data model metadata: Curation.* -->
  <PARAM ID="Reference" name="OUTPUT:Reference" datatype="char" ucd="meta.bib.bibcode"
      utype="ssa:Curation.Reference" arraysize="*" value="" >
      <DESCRIPTION>URL or Bibcode for documentation</DESCRIPTION>
  </PARAM>
  <PARAM ID="Publisher" datatype="char" name="OUTPUT:Publisher" ucd="meta.curation"
      utype="ssa:Curation.Publisher" value="ESO/VOS" arraysize="*" >
      <DESCRIPTION>Dataset publisher</DESCRIPTION>
  </PARAM>

  <!-- data model metadata: Target.* -->
  <PARAM ID="TargetName" name="OUTPUT:TargetName" datatype="char" ucd="meta.id;src"
      utype="ssa:Target.Name" arraysize="*" value="" >
      <DESCRIPTION>Target name</DESCRIPTION>
  </PARAM>

  <!-- data model metadata: CoordSys.* -->
  <PARAM ID="SpaceFrameName" name="OUTPUT:SpaceFrameName" datatype="char"
      utype="ssa:CoordSys.SpaceFrame.Name" arraysize="*" value="" >
      <DESCRIPTION>Spatial coordinate frame name</DESCRIPTION>
  </PARAM>
  <PARAM ID="SpaceFrameEquinox" name="OUTPUT:SpaceFrameEquinox" datatype="double"
      ucd="time.equinox;pos.frame" utype="ssa:CoordSys.SpaceFrame.Equinox"
      unit="yr" value="" >
      <DESCRIPTION>Equinox</DESCRIPTION>
  </PARAM>

<!-- characterization metadata -->

  <!-- characterization metadata: Char.FluxAxis -->
  <PARAM ID="FluxAxisUcd" name="OUTPUT:FluxAxisUcd" datatype="char"
      utype="ssa:Char.FluxAxis.Ucd" arraysize="*" value="" >
      <DESCRIPTION>UCD for flux</DESCRIPTION>
  </PARAM>

  <!-- characterization metadata: SpectralAxis -->
  <PARAM ID="SpectralAxisUcd" name="OUTPUT:SpectralAxisUcd" datatype="char"
      utype="ssa:Char.SpectralAxis.Ucd" arraysize="*" value="" >
      <DESCRIPTION>UCD for spectral coord</DESCRIPTION>
  </PARAM>
```

```
<!-- characterization metadata: Char.*.Coverage -->
<PARAM ID="TimeLocation" name="OUTPUT:TimeLocation" datatype="double"
    ucd="time.epoch" utype="ssa:Char.TimeAxis.Coverage.Location.Value"
    unit="d" value="" >
    <DESCRIPTION>Midpoint of exposure on MJD scale</DESCRIPTION>
</PARAM>
<PARAM ID="TimeExtent" name="OUTPUT:TimeExtent" datatype="double"
    ucd="time.duration;obs.exposure"
    utype="ssa:Char.TimeAxis.Coverage.Bounds.Extent" unit="s" value="" >
    <DESCRIPTION>Total exposure time</DESCRIPTION>
</PARAM>
<PARAM ID="TimeStart" name="OUTPUT:TimeStart" datatype="double"
    ucd="time.start;obs.exposure"
    utype="ssa:Char.TimeAxis.Coverage.Bounds.Start" unit="d" value="" >
    <DESCRIPTION>Start time</DESCRIPTION>
</PARAM>
<PARAM ID="TimeStop" name="OUTPUT:TimeStop" datatype="double"
    ucd="time.end;obs.exposure"
    utype="ssa:Char.TimeAxis.Coverage.Bounds.Stop" unit="d" value="" >
    <DESCRIPTION>Stop time</DESCRIPTION>
</PARAM>
<PARAM ID="SpatialLocation" name="OUTPUT:SpatialLocation" datatype="double"
    ucd="pos.eq" utype="ssa:Char.SpatialAxis.Coverage.Location.Value" arraysize="2"
    unit="deg" value="" >
    <DESCRIPTION>Spatial Position</DESCRIPTION>
</PARAM>

<!-- characterization metadata: Char.*.Accuracy -->
<PARAM ID="FluxCalibration" name="OUTPUT:FluxCalibration" datatype="char"
    utype="ssa:Char.FluxAxis.Accuracy.Calibration" arraysize="*" value="" >
    <DESCRIPTION>Type of flux calibration</DESCRIPTION>
</PARAM>
<PARAM ID="TimeCalibration" name="OUTPUT:TimeCalibration" datatype="char"
    ucd="meta.code.qual" utype="ssa:Char.TimeAxis.Accuracy.Calibration"
    arraysize="*" value="" >
    <DESCRIPTION>Type of coord calibration</DESCRIPTION>
</PARAM>
<PARAM ID="SpatialCalibration" name="OUTPUT:SpatialCalibration" datatype="char"
    ucd="meta.code.qual" utype="ssa:Char.SpatialAxis.Accuracy.Calibration"
    arraysize="*" value="" >
    <DESCRIPTION>Type of spatial coord calibration</DESCRIPTION>
</PARAM>

</RESOURCE>
</VOTABLE>
```

# Appendix D: SSA Data Model Summary

| UTYPE | UCD | Description | DataType | ArraySize |
|--------|-----|-------------|----------|-----------|
| | | | | |
| **Query** | | **Query Metadata** | | |
| Query.Score | | Degree of match to query parameters | float | |
| Query.Token | | Continuation token for large queries | char | * |
| | | | | |
| **Association** | | **Association Metadata** | | |
| Association.Type | | Type of association | char | * |
| Association.ID | | Association identifier | char | * |
| Association.Key | | Key used to distinguish association elements | char | * |
| | | | | |

| Access | | Access Metadata | | |
|---|---|---|---|---|
| Access.Reference | meta.ref.url | URL used to access dataset | char | * |
| Access.Format | | Content or MIME type of dataset | char | * |
| Access.Size | | Estimated dataset size | long | |
| | | | | |
| Protocol | | Protocol Metadata | | |
| ssa | | XML name space for SSA protocol | | |
| spec | | XML name space for Spectrum data model | | |
| | | | | |
| Spectrum | | General Dataset Metadata | | |
| Dataset.DataModel | | Datamodel name and version | char | * |
| Dataset.Type | | Dataset or segment type | char | * |
| Dataset.Length | meta.number | Number of points | long | |
| Dataset.Deleted | | Set if dataset is deleted | char | * |
| Dataset.TimeSI | | SI factor and dimensions | char | * |
| Dataset.SpectralSI | | SI factor and dimensions | char | * |
| Dataset.FluxSI | | SI factor and dimensions | char | * |
| Dataset.SpectralAxis | | Table column containing spectral coord | char | * |
| Dataset.FluxAxis | | Table column containing flux values | char | * |
| | | | | |
| DataID | | Dataset Identification Metadata | | |
| DataID.Title | meta.title;meta.dataset | Dataset Title | char | * |
| DataID.Creator | | Dataset creator | char | * |
| DataID.Collection | | Data collection to which dataset belongs | char | * |
| DataID.DatasetID | meta.id;meta.dataset | IVOA Dataset ID | char | * |
| DataID.CreatorDID | meta.id | Creator's ID for the dataset | char | * |
| DataID.Date | time;meta.dataset | Data processing/creation date | char | * |
| DataID.Version | meta.version;meta.dataset | Version of dataset | char | * |
| DataID.Instrument | meta.id;instr | Instrument name | char | * |
| DataID.Bandpass | instr.bandpass | Band as in RSM Coverage.Spectral | char | * |
| DataID.DataSource | | Original source of the data | char | * |
| DataID.CreationType | | Dataset creation type | char | * |
| DataID.Logo | meta.ref.url | URL for creator logo | char | * |
| DataID.Contributor | | Contributor | char | * |
| | | | | |
| Curation | | Curation Metadata | | |
| Curation.Publisher | meta.curation | Dataset publisher | char | * |
| Curation.PublisherID | meta.ref.url;meta.curation | URI for VO Publisher | char | * |
| Curation.PublisherDID | meta.ref.url;meta.curation | Publisher's ID for the dataset ID | char | * |
| Curation.Date | | Date curated dataset last modified | char | * |
| Curation.Version | meta.version;meta.curation | Publisher's version of the dataset | char | * |
| Curation.Rights | | Restrictions on data access | char | * |
| Curation.Reference | meta.bib.bibcode | URL or Bibcode for documentation | char | * |
| Curation.Contact.Name | meta.bib.author;meta.curation | Contact name | char | * |
| Curation.Contact.Email | meta.ref.url;meta.email | Contact email | char | * |
| | | | | |
| Target | | Target Metadata | | |
| Target.Name | meta.id;src | Target name | char | * |
| Target.Description | meta.note;src | Target description | char | * |

| Target.Class | src.class | Object class of observed target | char | * |
|---|---|---|---|---|
| Target.Pos | pos.eq;src | Target RA and Dec | double | 2 |
| Target.SpectralClass | src.spType | Object spectral class | char | * |
| Target.Redshift | src.redshift | Target redshift | double | |
| Target.VarAmpl | src.var.amplitude | Target variability amplitude (typical) | float | |
| | | | | |
| **Derived** | | **Derived Metadata** | | |
| Derived.SNR | stat.snr | Signal-to-noise for spectrum | float | |
| Derived.Redshift.Value | | Measured redshift for spectrum | double | |
| Derived.Redshift.StatError | stat.error;src.redshift | Error on measured redshift | float | |
| Derived.Redshift.Confidence | | Confidence value on redshift | float | |
| Derived.VarAmpl | src.var.amplitude;arith.ratio | Variability amplitude as fraction of mean | float | |
| | | | | |
| **CoordSys** | | **Coordinate System Metadata** | | |
| CoordSys.ID | | ID string for coordinate system | char | * |
| CoordSys.SpaceFrame.Name | | Spatial coordinate frame name | char | * |
| CoordSys.SpaceFrame.Ucd | meta.ucd | Space frame UCD | char | * |
| CoordSys.SpaceFrame.RefPos | | Origin of SpaceFrame | char | * |
| CoordSys.SpaceFrame.Equinox | time.equinox;pos.frame | Equinox | double | |
| CoordSys.TimeFrame.Name | time.scale | Timescale | char | * |
| CoordSys.TimeFrame.Ucd | meta.ucd | Time frame UCD | char | * |
| CoordSys.TimeFrame.Zero | arith.zp;time | Zero point of timescale in MJD | double | |
| CoordSys.TimeFrame.RefPos | time.scale | Location for times of photon arrival | char | * |
| CoordSys.SpectralFrame.Name | | Spectral frame name | char | * |
| CoordSys.SpectralFrame.Ucd | meta.ucd | Spectral frame UCD | char | * |
| CoordSys.SpectralFrame.RefPos | sdm:spect.frame | Spectral frame origin | char | * |
| CoordSys.SpectralFrame.Redshift | | Redshift value used if restframe corrected | double | |
| CoordSys.RedshiftFrame.Name | | Redshift frame name | char | * |
| CoordSys.RedshiftFrame.DopplerDefinition | | Type of redshift | char | * |
| CoordSys.RedshiftFrame.RefPos | | Redshift frame origin | char | * |
| | | | | |
| **Char.SpatialAxis** | | **Spatial Axis Characterization** | | |
| Char.SpatialAxis.Name | | Name for spatial axis | char | * |
| Char.SpatialAxis.Ucd | meta.ucd | UCD for spatial coord | char | * |
| Char.SpatialAxis.Unit | meta.unit | Unit for spatial coord | char | * |
| Char.SpatialAxis.Coverage.Location.Value | pos.eq | Spatial Position | double | 2 |
| Char.SpatialAxis.Coverage.Bounds.Extent | instr.fov | Aperture angular size | double | |
| Char.SpatialAxis.Coverage.Support.Area | | Aperture region | char | * |
| Char.SpatialAxis.Coverage.Support.Extent | instr.fov | Field of view area, sq. deg. | double | |
| Char.SpatialAxis.SamplingPrecision.SampleExtent | phys.angSize;instr.pixel | Spatial bin size | float | |
| Char.SpatialAxis.SamplingPrecision.FillFactor | stat.filling;pos.eq | Spatial sampling filling factor | float | |
| Char.SpatialAxis.Accuracy.StatError | stat.error;pos.eq | Astrometric statistical error | double | |
| Char.SpatialAxis.Accuracy.SysError | stat.error.sys;pos.eq | Astrometric systematic error | double | |
| Char.SpatialAxis.Calibration | meta.code.qual | Type of spatial coord calibration | char | * |
| Char.SpatialAxis.Resolution | pos.angResolution | Spatial resolution of data | double | |
| | | | | |
| **Char.SpectralAxis** | | **Spectral Axis Characterization** | | |
| Char.SpectralAxis.Name | | Name for spectral axis | char | * |
| Char.SpectralAxis.Ucd | meta.ucd | UCD for spectral coord | char | * |
| Char.SpectralAxis.Unit | meta.unit | Unit for spectral coord | char | * |
| Char.SpectralAxis.Coverage.Location.Value | em.wl;instr.bandpass | Spectral coord value | double | |
| Char.SpectralAxis.Coverage.Bounds.Extent | em.wl;instr.bandwidth | Width of spectrum | double | |
| Char.SpectralAxis.Coverage.Bounds.Start | em.wl;stat.min | Start in spectral coordinate | double | |
| Char.SpectralAxis.Coverage.Bounds.Stop | em.wl;stat.max | Stop in spectral coordinate | double | |
| Char.SpectralAxis.Coverage.Support.Extent | em.wl;instr.bandwidth | Effective width of spectrum | double | |
| Char.SpectralAxis.SamplingPrecision.SampleExtent | em.wl;spect.binSize | Wavelength bin size | double | |
| Char.SpectralAxis.SamplingPrecision.FillFactor | stat.filling;em | Spectral sampling filling factor | float | |
| Char.SpectralAxis.Accuracy.BinSize | em.wl;spect.binSize | Spectral coord bin size | double | |
| Char.SpectralAxis.Accuracy.StatError | stat.error;em | Spectral coord statistical error | double | |
| Char.SpectralAxis.Accuracy.SysError | stat.error.sys;em | Spectral coord systematic error | double | |
| Char.SpectralAxis.Calibration | meta.code.qual | Type of spectral coord calibration | char | * |
| Char.SpectralAxis.Resolution | spect.resolution;em | Spectral resolution FWHM | double | |
| Char.SpectralAxis.ResPower | spect.resolution | Spectral resolving power | float | |
| | | | | |
| **Char.TimeAxis** | | **Time Axis Characterization** | | |
| Char.TimeAxis.Name | | Name for time axis | char | * |
| Char.TimeAxis.Ucd | meta.ucd | UCD for time | char | * |
| Char.TimeAxis.Unit | meta.unit | Unit for time | char | * |
| Char.TimeAxis.Coverage.Location.Value | time.epoch | Midpoint of exposure on MJD scale | double | |
| Char.TimeAxis.Coverage.Bounds.Extent | time.duration | Total exposure time | double | |
| Char.TimeAxis.Coverage.Bounds.Start | time.start;obs.exposure | Start time | double | |
| Char.TimeAxis.Coverage.Bounds.Stop | time.end;obs.exposure | Stop time | double | |
| Char.TimeAxis.Coverage.Support.Extent | time.duration;obs.exposure | Effective exposure time | double | |
| Char.TimeAxis.SamplingPrecision.SampleExtent | time.interval | Time bin size | double | |
| Char.TimeAxis.SamplingPrecision.FillFactor | stat.filling;time | Time sampling filling factor | float | |
| Char.TimeAxis.Accuracy.BinSize | time.interval | Time bin size | double | |
| Char.TimeAxis.Accuracy.StatError | stat.error;time | Time coord statistical error | double | |
| Char.TimeAxis.Accuracy.SysError | stat.error.sys;time | Time coord systematic error | double | |
| Char.TimeAxis.Calibration | meta.code.qual | Type of coord calibration | char | * |
| Char.TimeAxis.Resolution | time.resolution | Temporal resolution FWHM | double | |
| | | | | |
| **Char.FluxAxis** | | **Flux Axis Characterization** | | |
| Char.FluxAxis.Name | | Name for flux | char | * |
| Char.FluxAxis.Ucd | meta.ucd | UCD for flux | char | * |
| Char.FluxAxis.Unit | meta.unit | Unit for flux | char | * |
| Char.FluxAxis.Accuracy.StatError | stat.error;phot.flux.density;em | Flux statistical error | double | |
| Char.FluxAxis.Accuracy.SysError | stat.error.sys;phot.flux.density;em | Flux systematic error | double | |
| Char.FluxAxis.Calibration | | Type of flux calibration | char | * |